\documentclass{article}
 \pdfoutput=1 

\usepackage{arxiv}

\usepackage[utf8]{inputenc} 
\usepackage[T1]{fontenc}    
\usepackage{hyperref}       
\usepackage{url}            
\usepackage{booktabs}       
\usepackage{amsfonts}       
\usepackage{nicefrac}       
\usepackage{microtype}      
\usepackage{lipsum}

\usepackage{todonotes}
\usepackage{amsmath}
\usepackage{afterpage}
\usepackage{appendix}

\usepackage{graphicx}
\usepackage{dcolumn}
\usepackage{bm}
\usepackage{epstopdf}

\usepackage[utf8]{inputenc}
\usepackage[T1]{fontenc}
\usepackage{mathptmx}

\title{Timing and Shape of Stochastic Autocatalytic Burst Formation}

\author{
  Alastair Jamieson-Lane \\
 Department of Mathematics,\\ University of British Columbia\\
  \texttt{aja107@math.ubc.ca} \\
   \And
 Eric N. Cytrynbaum \\
  Department of Mathematics,\\ University of British Columbia.\\
  \texttt{cytryn@math.ubc.ca} \\
}

\begin{document}
\maketitle

\begin{abstract}
Chemical, physical and ecological systems passing through a saddle-node bifurcation will, momentarily, find themselves balanced at a semi-stable steady state. 
If perturbed by noise, such systems will escape from the zero-steady state, with escape time sensitive to noise.  
When the model is extended to include space, this leads to different points in space ``escaping from zero'' at different times, and uniform initial conditions nucleate into sharp peaks spread randomly across a nearly uniform background, a phenomena closely resembling nucleation during phase transition. We use Large Deviation Theory to determine burst shape and temporal scaling with respect to noise amplitude. These results give a prototype for a particular form of patternless symmetry breaking in the vicinity of a stability boundary, and demonstrates how microscopic noise can lead to macroscopic effects in such a region. 
\end{abstract}


\maketitle

\section{Introduction}

The universe we live in does not consist of a perfectly uniform cloud of hydrogen gas, nor is the earth a perfect sphere. Clouds condense into droplets of rain \cite{mcgraw_kinetic_2003}, a single zygote divides and differentiates into a multitude of different cell types, a flipped coin lands either heads, or tails.
All around us, symmetry is broken every day.

Depending on the system of interest, symmetry can be broken in either a patterned or unpatterned manner. Pattern formation is ubiquitous in nature, from a Zebra's strips \cite{turing_chemical_1952}  to Tiger bush \cite{lefever_origin_1997}. Patterned symmetry breaking in reaction diffusion systems is well understood, and was first studied by Alan Turing in the 1950s \cite{turing_chemical_1952}.

In nature we also observe patternless symmetry breaking, in for example the nucleation of crystals\cite{erdemir_nucleation_2009} or rain drops\cite{mcgraw_kinetic_2003}.
 This patternless symmetry breaking is less studied in a mathematical context, and is the focus of this article.

\begin{figure*}
\includegraphics[width=0.95\textwidth]{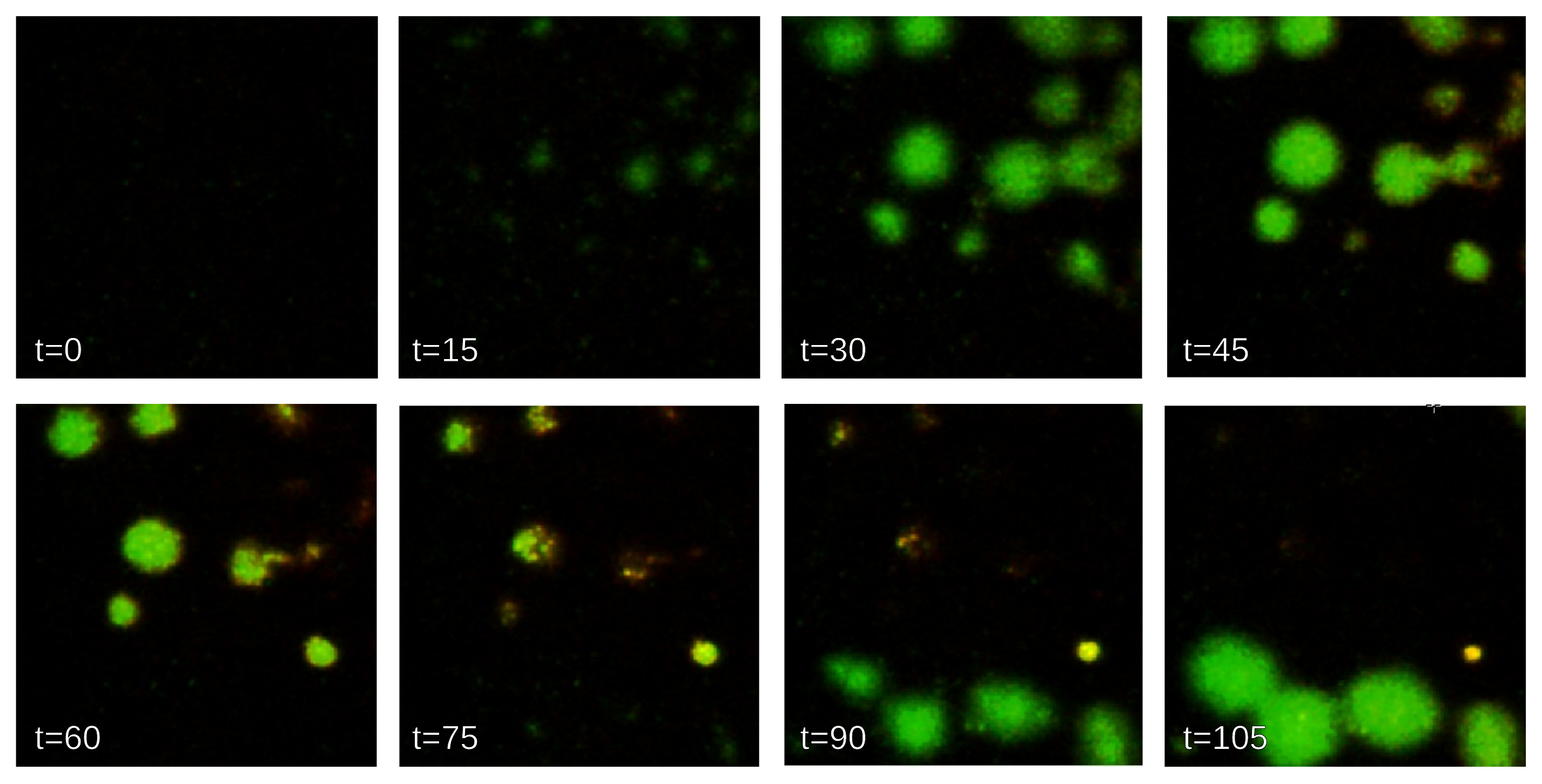}
\caption{Burst patterning, as observed by Vecchiarelli \emph{et al}. At $t=0$, the membrane is at a low, almost homogeneous, concentration of particles. This initial condition evolves into bright, high concentration `bursts' of membrane-bound MinD, which then freeze in place, and fade. While one generation of bursts fades, the next generation of bursts appears (bottom row). These previously unpublished images of the experiments described in \cite{vecchiarelli_membrane-bound_2016} were kindly provided by A. Vecchiarelli.
}
\label{fig:burstPicture}
    \end{figure*}

Our present work was motivated by intriguing experimental results in a recent study of the Min cell division system in E.coli \cite{vecchiarelli_membrane-bound_2016}.  In their study, Vecchiarelli et al. observed that Min proteins demonstrate either patterned or unpatterned symmetry breaking, depending on the concentration conditions. For high concentrations of the relevant Min proteins, spirals and linear waves are observed, with each of these patterns possessing their own well studied mathematical frameworks \cite{murray_mathematical_2008}. For lower concentrations, Vecchiarelli et al's experiment also demonstrates ``burst'' behavior, whereby the local concentration of Min proteins increases for spatially localized patches of membrane, and then reduces again shortly thereafter (see figure \ref{fig:burstPicture}). These bursts appear to arise from a relatively homogeneous background, and in contrast to the ordered high concentration patterns described above, they are positioned randomly across the membrane, and appear governed by low level particle scale noise.

In this paper, we demonstrate how dynamical systems in the vicinity of a semi-stable steady state are sensitive to low level white noise, giving rise to similar burst formation behavior.
We consider a ``canonical'' stochastic PDE model of such a system
\begin{equation}
    u_t = u_{xx} + u^2 + \epsilon \xi,
    \label{eq:canonModel}
\end{equation}
and use Large Deviation Theory to predict both the time of burst formation, and the shape of the burst formed, in the limit of low amplitude noise.

We consider a particular reaction-diffusion system from the literature in Sect. \ref{sect:simulations}, and use simulations  to demonstrate the occurrence of burst formation in appropriate parameter regimes. 
In Sec. \ref{sect:AnalyticsSpaceless}, we use Large Deviation Theory to predict the time taken for the stochastic ODE  $u_t=u^2 +\epsilon \xi$ to evolve from $u=0$ to some fixed value $U_f$, and determine the most probable path for this occurrence.
Sect. \ref{sect:AnalyticsSpacey} extends this work to systems with one spatial dimension via the addition of a diffusion term (as is \ref{eq:canonModel}). We use a mixture of numeric and analytic techniques to determine both the shape and time taken for bursts to form.
In Sec. \ref{sect:compResult}, we compare our analytic results both to one another, and to the corresponding simulations, and in Sec. \ref{sect:twoDbad} we discuss the conceptual difficulties preventing our analysis from being extended to higher dimensions.
Finally, in Sec. \ref{sect:Conclude} we summarize the results, and discuss limitations, physical relevance, and future directions for research. The Appendices contain both an introduction to Large Deviation Theory for those unfamiliar with the topic, along with more detailed calculations excluded from the main discussion.

\section{Computational Exploration}
\label{sect:simulations}
Before delving into analytical results, we take a brief computational detour to demonstrate the general behaviour of the class of neutrally stable systems that form the focus of this paper.\\
\begin{figure*}
\begin{minipage}[b]{\textwidth}
\includegraphics[width=1.075\textwidth, trim= 100 0 00 0]{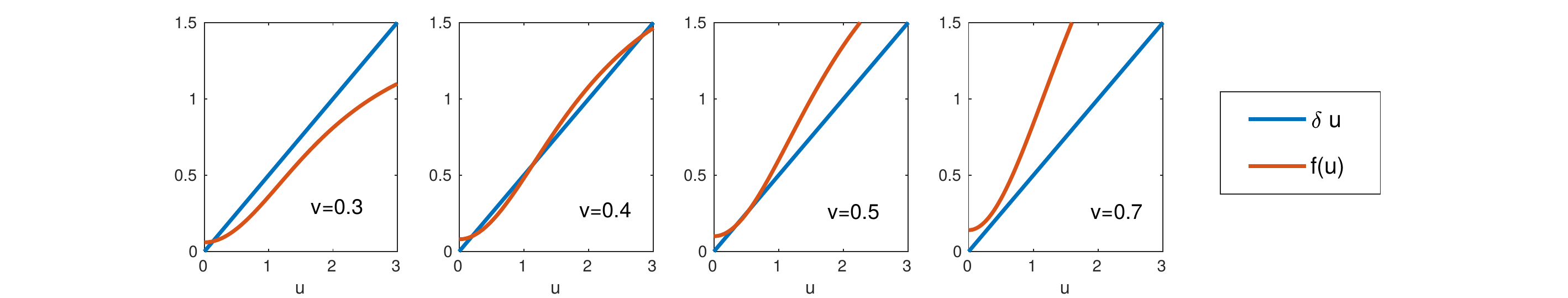}
\caption{As the bulk concentration $v$ is varied, $u_t= f(u,v)-\delta u= v \left(k_0 + \frac{\gamma u^2}{K^2+u^2} \right)- \delta u$ gains and loses steady states via saddle-node bifurcations. In particular, for $v=0.3$, only the low $u$ steady state exists. At $v=0.4$ there are three steady states- stable steady states near $u=0$ and $u=3$, with an unstable steady state. At $v\approx0.5$ the low $u$ steady state is lost, and only the high $u$ steady state remains (outside the bounds of these diagrams)}
\label{fig:MoriPhaseLine}
\includegraphics[width=0.975\textwidth]{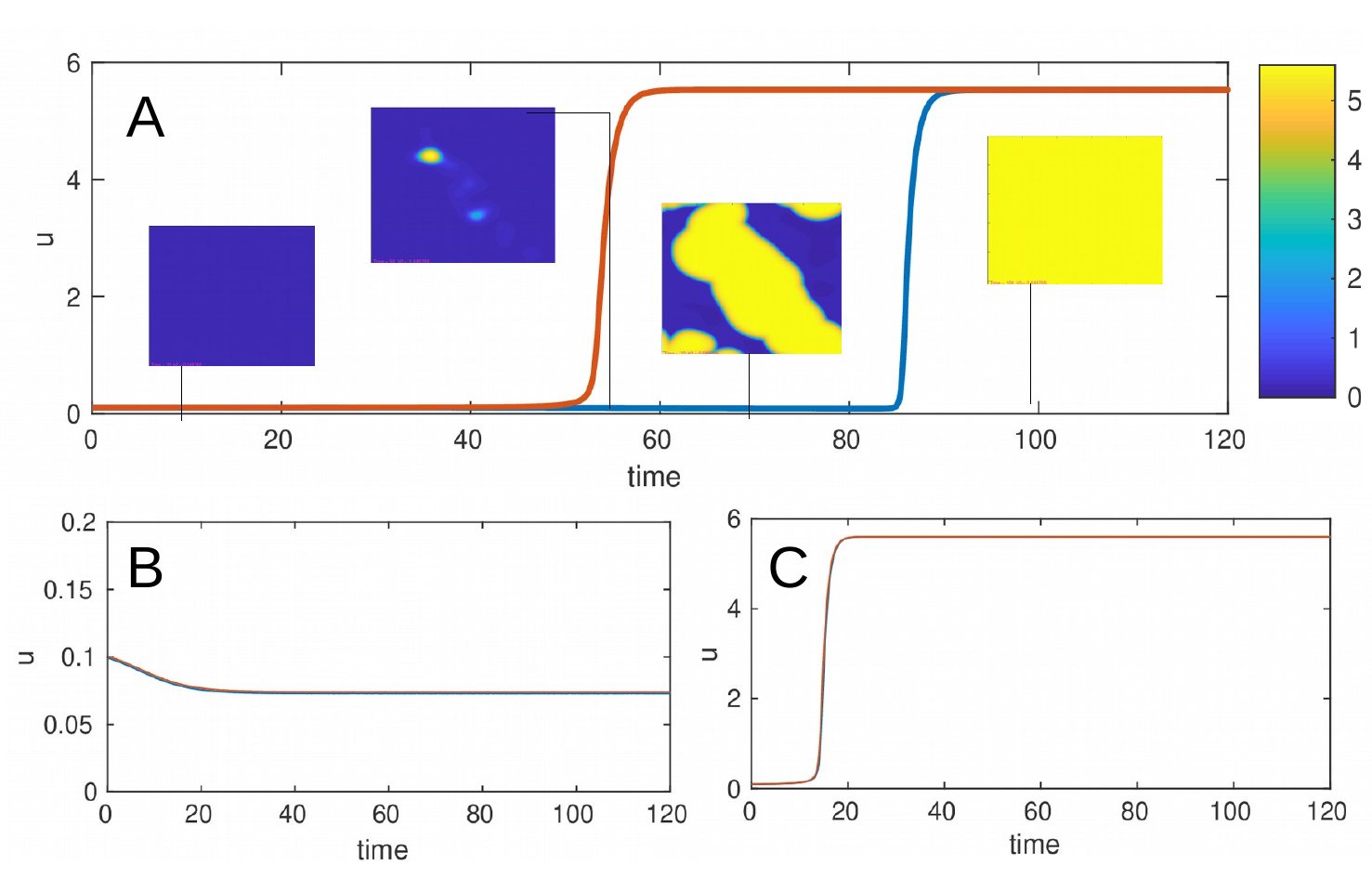}
\caption{For equation \protect\eqref{eq:moriBig} we track the maximum and minimum value of $u$ through time (red and blue, respectively). 
A) For $v$ selected at the bifurcation point, noise leads to burst formation (see images, inset), leading $\max(u)$ and $\min(u)$ to separate for a time. B) for $v$ selected slightly below the bifurcation value, noise is unable to overcome the energy barrier, and $u$ remains in the vicinity of the low $u$ steady state indefinitely. C) For $v$ slightly above the bifurcation value, $u$ tends towards saturation, but does so homogeneously - $\max(u)$ and $\min(u)$ remain close together throughout the entire process.}
\label{fig:VicinityOfBifurcation}
\end{minipage}
 \end{figure*}


Consider the following equation due to Mori et al. \cite{mori_wave-pinning_2008}
\begin{align}
u_t = D \Delta u + f(u,v) - \delta u \label{eq:moriBig} \\ 
\text{where } f(u,v) = v \left(k_0 + \frac{\gamma u^2}{K^2+u^2} \right) ,
\end{align}
over the unit square in $\mathbb{R}^2$ with periodic boundary conditions. For the sake of these simulations, we use $\gamma=5$, $K=2$ $\delta=0.5$ and diffusion coefficient $D=10^{-4}$.

Equation \eqref{eq:moriBig} was designed to model a system of cell signaling proteins called Rho GTPases. The model was originally used to demonstrate wave pinning, a behavior reminiscent of the ``cell polarization'' behavior associated with Rho proteins.

In this equation $D$, $k_0$, $\gamma$, $K$ and $\delta$ are reaction and diffusion parameters, $u$ is the concentration of membrane bound Rho protein and $v$ is the concentration of Rho diffusing quickly through the bulk of the cell.
As $v$ is varied, the above system gains and loses steady states via a pair of saddle-node bifurcations (see figure \ref{fig:MoriPhaseLine}). 
Let $u_0$ and $v_0$ denote the values of $u$ and $v$ at the second of these saddle-node bifurcations, in which the lower stable steady state is annihilated. In this region $v_0 \approx 0.5$.

Simulating with fixed $v=v_0$, $u(0,{\bf x})=u_0$ and additive noise of the form $\epsilon \xi$, leads to a long nearly homogeneous period where $u({\bf x},t) \approx u_0$ for all ${\bf x}$, followed by the abrupt formation of several ``spikes'', which then saturate and spread across the domain. (See figure \ref{fig:VicinityOfBifurcation} A)

By contrast, when we select some fixed $v$ greater than $v_0$ the system remains approximately homogeneous as it tend towards this $u$-saturated state. For $v$ less than $v_0$, the system instead approaches some homogeneous low $u$ state. In either of these latter two cases, no significant patterning or heterogeneity is observed. (See figure \ref{fig:VicinityOfBifurcation} B \& C)

Similar results may be observed for a variety of spatially distributed systems in the vicinity of saddle-node bifurcations. In order to consider this behavior in more generality, we consider a variety of ``cannonical'' equations, each demonstrating different aspects of the behavior of interest in the vicinity of a generic saddle-node.
    
\section{Escape time from steady states; spaceless case}
\label{sect:AnalyticsSpaceless}
Let us begin our discussion by considering first the spaceless stochastic differential equation
\begin{align}
\label{eq:SDE_basic}
\dot u= u^2 + \epsilon \xi.
\end{align}
Here $u$ is our stochastic process of interest, $\epsilon$ is some small noise amplitude parameter, and $\xi$ is assumed to be Gaussian white noise. We assume noise as defined by Walsh \cite{walsh_introduction_1986}, such that the integral over any time interval $\Omega$ is given by $\int_\Omega \xi dt = N(0,|\Omega|)$, and integrals over non-intersecting time intervals are independent. The extension to spatio-temporal white noise (also discussed by Walsh) makes use of higher dimensional integration. Equation \ref{eq:SDE_basic} provides a canonical example of a spaceless noisy system balanced at a saddle node, and provides a toy model on which to build our understanding before moving to the spatial case.

Deterministically, trajectories starting at $u=0$ remain stationary, and never reach positive $U_f$. However, any noise added to the system is enough to disturb this equilibrium and allow $u$ to escape.
We would like to determine the amount of time it takes our stochastic process $u$ to travel from the semi-stable steady state $u=U_i=0$ to some positive constant $U_f$. The appropriate tool for this task is Large Deviation Theory.

Large Deviation Theory (LDT) is a theory used to study unlikely noise driven events in stochastic systems (often, but not always SDEs).
For the interested reader, we give a brief introduction to LDT in Appendix \ref{ap:IntroToLDT}. For those who wish for more detail, we recommend Freidln and Wentzell's ``Random Perturbations of Dynamical Systems'' \cite{freidlin_random_2012}, and Rassoul-agha and Seppalainen's  ``A Course on Large Deviations With an Introduction to Gibbs Measures''\cite{rassoul-agha_course_2015}. Here we present only the key results of LDT - those concepts and theorems critical to our present interest.

The central premise of LDT is that when unlikely events occur, they are overwhelmingly likely to occur via a path “close to” the most probable path. The most probable path, $\phi(t)$, is found by minimizing the ``normalized action functional''
\begin{align}
\bar S_T(\xi) = \int_0^T  \epsilon^2 \xi ^2 dt.
\label{eq:SdefineXi}
\end{align}

For any given SPDE, $\bar S_T(\xi)$ can be re-written as a function of $\phi$. For example, when studying eq. \ref{eq:SDE_basic} we can rearrange to isolate $\epsilon \xi$ and find
\begin{align}
 S_T(\phi) = \int_0^T  (\dot \phi - \phi^2)^2 dt
\label{eq:SdefinePhi}
\end{align}

Minimizing $S_T(\phi)$  allows us to determine the most probably path for a given event, and having identified this part we are able to approximate the probability of an event occurring by a given time using either
\begin{equation}
\label{eq:LDT2}
\lim_{\epsilon\rightarrow 0} \epsilon^2 \ln \mathbb{E}(\tau)= \min_{T,\phi} \bar S_T(\phi),
\end{equation}
to estimate the expected time $\mathbb{E}(\tau)$ or 
\begin{equation}
\lim_{\epsilon \rightarrow 0} \epsilon^2 \ln P(\tau \le T) = - \min_\phi \Bar{S}_T(\phi)
\label{eq:LDTmain}
\end{equation}
to estiamte the probability distribution.

These formula are given as theorem 4.1 and theorem 1.2 (respectively) in chapter 4 of  Freidln and Wentzell\cite{freidlin_random_2012}. 

We will make use of both of these theorems throughout this paper, however, because $\min_{T,\phi} S_T(\phi)=0$ for many of the examples discussed, eq. \ref{eq:LDTmain} will prove to be the more helpful of the two. This is because we are in the slightly unusual position of studying escape from an energy plateau, as opposed to an energy well, as might be more usually studied.

Suppose we wish to study the probability of $u$ passing from $U_i=0$ to $u\ge U_f$ by time $t=T$, that is to say the ``escape from zero'' problem for equation \ref{eq:SDE_basic}. In order to use either of the above formula, we must find the minimum of ${S}_T(\phi)$.
By the calculus of variations, any minimizer of ${S}_T(\phi)$ must satisfy 
\begin{align}
0=\frac{d \Bar{S}_T(\phi)}{d\phi}= -2 \ddot \phi + 4 \phi^3, 
\label{eq:MinimizationDefine}
\end{align}
and hence 
\begin{align}
\int_{U_i}^{\phi(t)} \frac{1}{\sqrt{\psi^4 +C} } d\psi= T-0.
\label{eq:Cpinning}
\end{align}
For $C=0$ we recover the deterministic solution $\phi(t) = 1/(U_i^{-1} -t)$. This solution indicates infinite travel time to $U_f$ in the case where $U_i=0$ (our case of interest).
For $C \neq 0$, the integral can be solved using wolfram\textbar alpha \cite{wolfram|alpha_wolfram|alpha:_2018}, determining $t$ in terms of $\phi$. Given that the function itself provides limited illumination, and is defined in terms of the ellipticF function (itself defined in terms of an integral), we will refer to solutions of \ref{eq:Cpinning}  simple as $\tau(\phi)$, the inverse of $\phi(t)$.

In order to match our assumed boundary conditions, we must pick $C$ such that $\tau(U_f)=T$. This is not generically an easy problem to solve exactly, but for large $U_f$ it can be well approximated.
In order to determine $C$, in this limit, we first note $\tau_\infty= \lim_{\phi \rightarrow \infty} \tau(\phi)= \frac{4 \Gamma(5/4)^2}{\sqrt{\pi} C^{1/4}}$ (as given by wolfram\textbar alpha \cite{wolfram|alpha_wolfram|alpha:_2018-1}). Here $\Gamma$ is the Gamma function \cite{artin_gamma_2015}, and takes the value $\Gamma(5/4) \approx 0.9064$. The deterministic time taken to get from $\phi= U_f$ to $\phi=\infty$ is $1/U_f$, thus $\tau_\infty \approx T + 1/U_f$. Rearranging gives 
\begin{align}
\label{eq:Capprox}
C \approx \left( \frac{4 \Gamma(5/4)^2}{\sqrt{\pi} (T + 1/U_f)} \right)^4.
\end{align}

Numerical experimentation indicates that, for this choice of $C$,  $|\tau(U_f) - T|/T<10^{-6}$ whenever $T,U_f \ge 4$, indicating that our approximation of $C$ is very good.

\begin{figure}[h]
    \centering
\includegraphics[width=0.435\columnwidth]{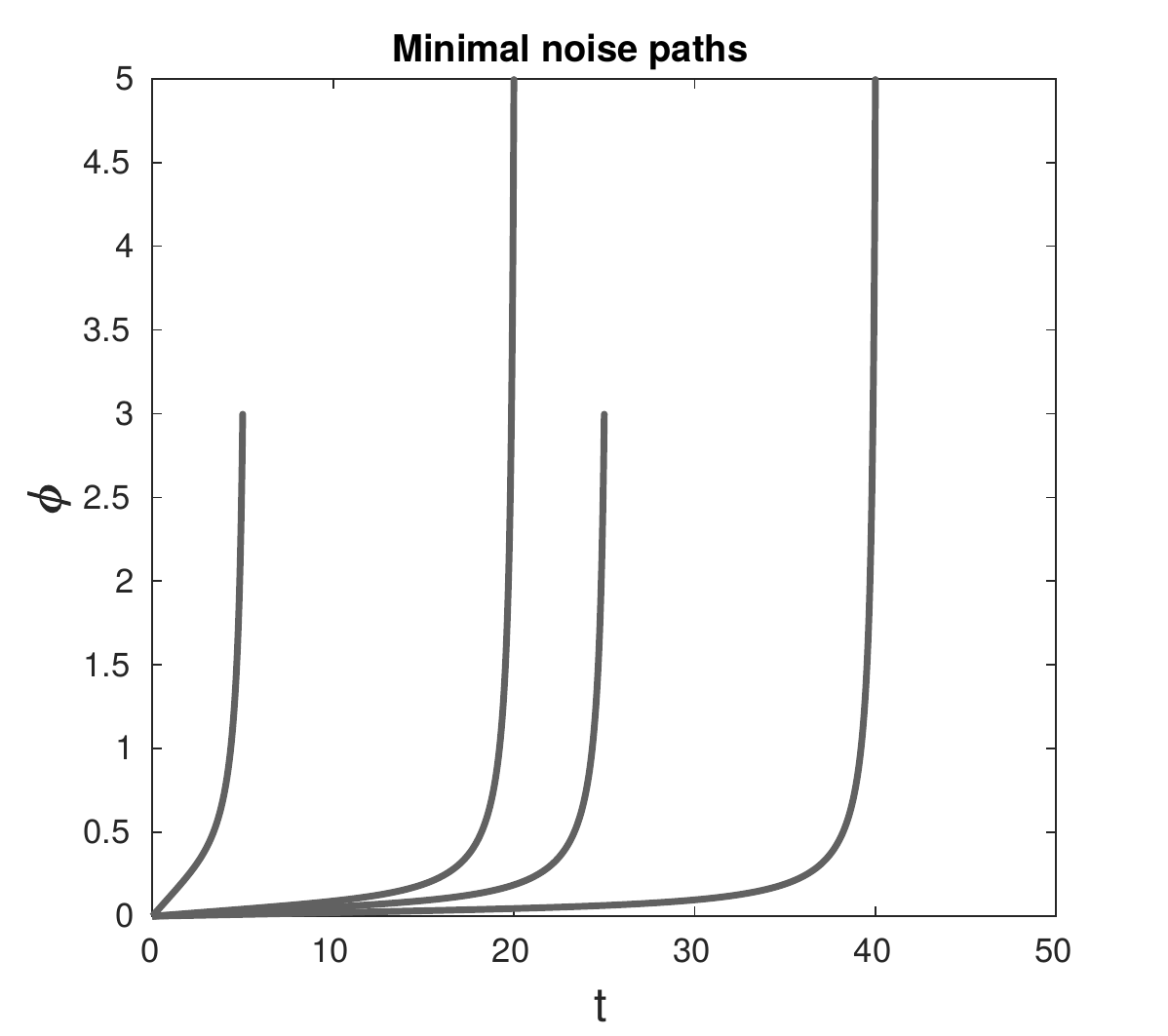}
\includegraphics[width=0.515\columnwidth]{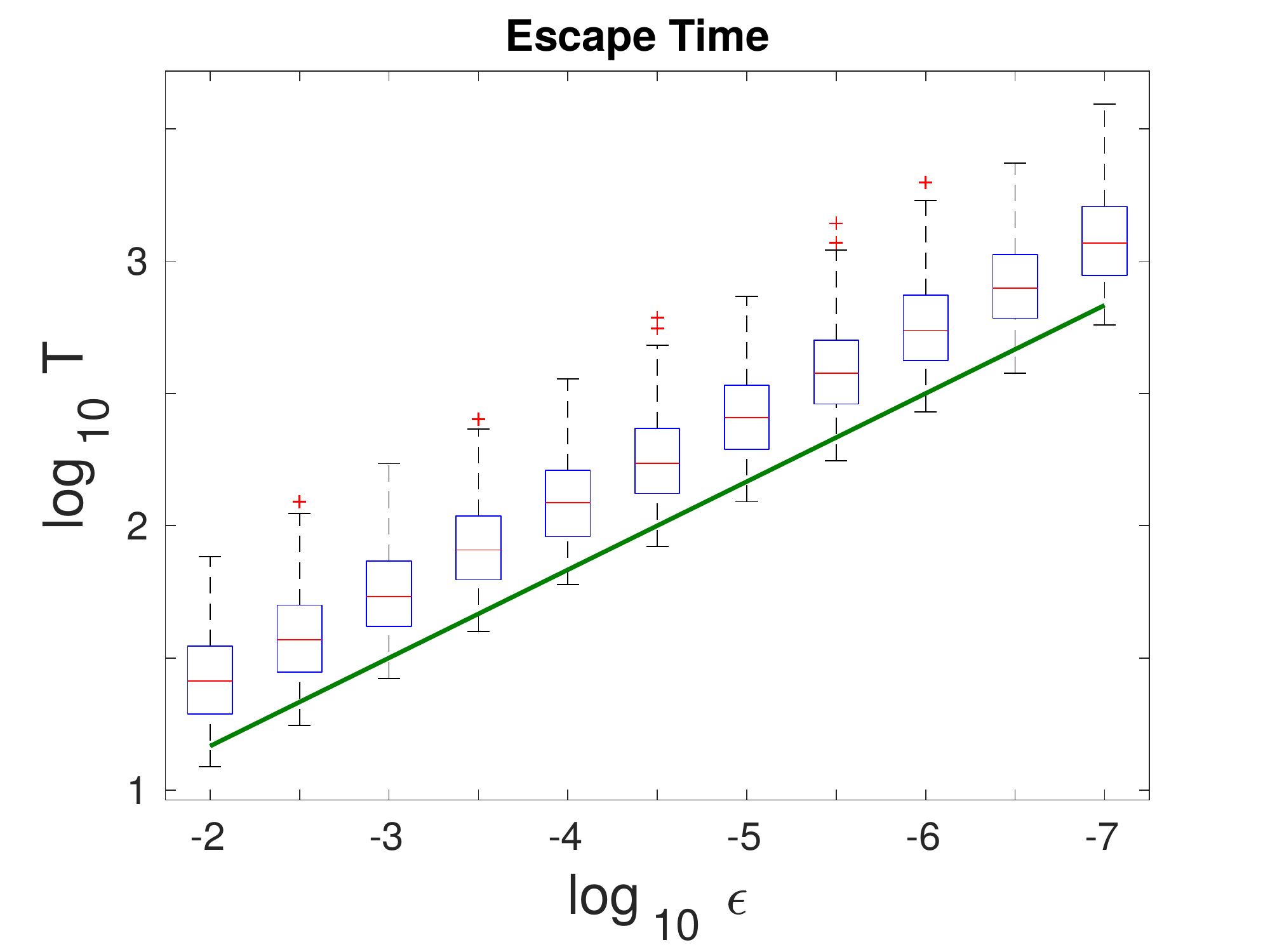}
\caption{(Left) Some optimal $\phi$ found for four particular boundary conditions, $\tau(U_f)=T$, using $\tau(\phi)$, as given by wolfram\textbar alpha \cite{wolfram|alpha_wolfram|alpha:_2018}. In all cases, $\tau(U_f)$ matches the requested $T$ to an exceptionally high degree of accuracy.
(Right) Passage time from $U_i=0$ to $U_f=1$ vs $\epsilon$ for $u_t = u^2 + \epsilon \xi$ (equation \ref{eq:SDE_basic}). For each $\epsilon$ value we use 4000 simulations. Boxplots indicate distribution of passage times. The solid green line indicates the predicted scaling of passage time $T= O(\epsilon^{-2/3})$. The slope from our theory matches simulation results. The vertical shift between theory and simulation is a reflection of the unknown coefficient in the scaling relationship.
}
\label{fig:PhiPathExamples}
    \end{figure}

Having determined $C$, we are able to plot $t=\tau(\phi)$ for arbitrary $U_f$ and $T$ (see figure 
\ref{fig:PhiPathExamples}, Left).

In order to invoke either eq. \ref{eq:LDT2} or eq. \ref{eq:LDTmain}, we must must determine $S_T(\phi)$ along this action minimizing path.
Starting from the definition of $S_T(\phi)$ (equation \ref{eq:SdefinePhi}), it is possible to show that in our case
\begin{align} 
{S}_T(\phi)
&= \frac{2}{3}\left[  \phi^3 ( \sqrt{1 +C \phi^{-4}} -1)\right]_0^{U_f} +  \frac{1}{3}C T,\\
  &\approx \frac{C}{3 U_f} - 0 +  \frac{1}{3}C T \text{ , \quad for large $U_f$}.
\label{eq:FinalSpacelessAction}
\end{align}
Details are given in appendix \ref{app:CunningByParts}. Remembering from equation (\ref{eq:Capprox}) that $C = O(T^{-4})$ we find ${S}_T(\phi) = O(T^{-3})$.

This analytic result agrees with numeric results found by approximating ${S}_T(\phi)$ directly for a variety of different $T$ values. 

Because ${S}_T(\phi) \rightarrow 0$ as $T \rightarrow \infty$, we see that eq. \ref{eq:LDT2} gives no information. This is a reflection of the fact that $\phi$ need not overcome any energy \emph{barrier} in its path from $0$ to $U_f$, but merely an energy plateau.  

We can however apply equation \ref{eq:LDTmain} to find
\begin{align}
\lim_{\epsilon\rightarrow 0} \epsilon^2 \ln[P(\tau \le T)] &= -\min_\phi \Bar{S}_T(\phi)= \frac{1}{3}CT, \\
\Rightarrow P(\tau \le T) &\sim \exp \left[ -k T^{-3} \epsilon^{-2} \right], 
\end{align}
for small $\epsilon$.
\begin{align}
\mathbb{E}(\tau) &= \int_0^\infty P(\tau > T) dt \approx \int_0^\infty 1- \exp \left[- k T^{-3} \epsilon^{-2} \right] dT,\\
&= O(\epsilon^{-2/3}) 
\end{align}

 Hence we find that time taken for a solution to $u_t= u^2 + \epsilon \xi$ to escape from zero and approach infinity is predicted to scale like $O(\epsilon^{-2/3})$. The majority of this time is spent close to $u=0$, and the escape time behavior of the system is dominated by the behavior of the system in this region; the exact value of $U_f$ has negligible impact, so long as $U_f$ is selected such that $U_f \gg O(\epsilon^{2/3})$. Simulation of equation \ref{eq:SDE_basic} for a variety of $\epsilon$ agrees with these asymptotic results (see figure \ref{fig:PhiPathExamples}, Right).

\subsection{Systems near a saddle-node bifurcation}
The above work describes the behaviour of the system precisely balanced \emph{at} a saddle-node bifurcation. For the sake of completeness, we might also consider systems in the vicinity of such a bifurcation.

The system 
\begin{align}
\label{eq:SDE_basic_unstable}
\dot u= u^2 + \gamma^2 + \epsilon \xi
\end{align}
is a canonical example of a system which, for small $\gamma$, has just lost its steady states. In the deterministic limit $\epsilon \rightarrow 0$, this system admits solutions of the form $T=\left[\tan^{-1}(u/\gamma)/\gamma \right]_{U_i}^{U_f}$, which gives travel times of order $O(\gamma^{-1})$. 

The system 
\begin{align}
\label{eq:SDE_basic_stable}
\dot u= u^2 - \gamma^2 + \epsilon \xi
\end{align}
is a canonical example of a system with a stable/unstable pair of steady states. Assuming in this case that $U_i=-\gamma$ (the stable steady state), and $U_f>\gamma$, it is possible to show that the action minimizing path obeys $\dot \phi^2 = (\phi^2 - \gamma^2)^2 +C^2$. In the limit of large $T$, we can select $C\rightarrow0$, and find $\Bar{S}_T(\phi) \rightarrow 16 \gamma^3/3$. 

By eq. \ref{eq:LDT2} we thus have $\mathbb{E}(\tau) \sim \exp[\epsilon^{-2} 16 \gamma^3/3]$.

In addition to the above static results, the behavior of systems passing through saddle node bifurcations (either with or without noise), has also been studied. A review of this ``delayed bifurcation theory'' is given by Christian Kuehn
\cite{kuehn_mathematical_2011}. For a more detailed introduction, see  Berglund and Gentz
\cite{berglund_noise-induced_2006}. We now move on to consider the previously unexplored spatially distributed problem.

\section{The Spatially Distributed Problem}
\label{sect:AnalyticsSpacey}
Now that we have built up our understanding using the spaceless model, let us turn our attention to the spatially distributed model.
Our primary interest in the spatially distributed case is considering the behavior of systems at a saddle-node bifurcation.
Spatial systems of this form can be represented via the canonical equation
\begin{equation}
  u_t= u_{xx} + u^2 + \epsilon \xi.
  \label{eq:burstWithNoise}
\end{equation}
Here, both $u$ and $\xi$ are functions of $x$ and $t$. We assume that $u$ starts at the steady state; $u(x,0)=0$. Here $\xi$ represents our white noise term, although it may be thought of as a forcing function, with an associated energy $\int \xi^2 dx dt$.

We wish to describe the behavior of the system as it `escapes' from the steady state at zero. We assume (without loss of generality) that $u$ takes its maximum at $x=0$, and ask ``for a given time $T$, what is the most probable path such that $u(0,T)=U_f$, where $u(x,t)<U_f$ for all $t<T$?'' 

For the sake of comparison, we will also consider escape from a linear unstable stead state,
\begin{equation}
  u_t= u_{xx} + u + \epsilon \xi.
  \label{eq:linearBurstWithNoise}
\end{equation}
While less mathematically interesting, this linear escape problem gives us an analytically tractable case to explore, along with something to compare our non-linear results to. We begin with this simpler case.

\subsection{Linearly Unstable Case}
\label{subsect:LinUnstable}
In what follows we present only the most important results, and the conceptually important steps leading to these results. Details can be found in Appendix \ref{Ap:AlgebraDetail}. 
In order to understand spike formation in equation
 (\ref{eq:linearBurstWithNoise}), we must minimize the associated ``action functional'':
\begin{equation}
S_T(u) = {\int_0}^{T} {\int_{-\infty}}^\infty (u_t-u_{xx}-u)^2 dx dt,
\label{eq:StartAnalysisLinear}
\end{equation}
subject to the conditions $u(x,0)=0$ and $u(0,T)=U_f$.
As per standard functional analysis techniques, this minima can be found when the functional derivative $\frac{d S_T(u)}{du}(x,t)=0$. This condition \emph{can} be written as a single differential equation with double the number of derivatives with respect to each variable, but is more conveniently written as a coupled \emph{pair} of PDEs:
\begin{equation}
 \begin{split}
 \label{eqn:backForwardBurstLinear}
  u_t= u_{xx} + u + \epsilon \xi, \\
  u(x,0)=0, \\
  -\xi_t= \xi_{xx} + \xi, \\
  \xi(x,T)= \alpha \delta(x).
\end{split}
\end{equation}
This system admits the explicit solution:
 \begin{align}
  \xi(x,t)&= \alpha e^{T-t} N(x,T-t),\\ 
  u(x,t)&= \int_0^t \alpha \epsilon e^{T+t-2 \tau} N(x,T+t-2 \tau) d\tau,
  \label{eq:bestLinearBurst}
  \end{align}
where 
 $
  N(x,\gamma)=  \frac{1}{\sqrt{4 \pi \gamma}} \exp \left[\frac{-x^2}{4\gamma} \right],
$
and $\alpha = \frac{4 U_f}{\epsilon erfi(\sqrt{2T})}$.

The associated action functional can be shown to be $S_T(u)= 4 U_f^2/erfi(2T) = U_f^2 O(e^{-2T})$.

Further analysis and discussion of these results we postpone until section \ref{sect:compSpaceResult}.

\subsection{Semi-stable Case}
We now consider the non-linear case, (equation \ref{eq:burstWithNoise}).
As previously, in order to determine the dynamics of spike formation, we concern ourselves primarily with determining the most probable path to spike formation. We present here only the conceptually important milestones in our calculations. Details are similar to those given in Appendix \ref{Ap:AlgebraDetail}.
\begin{equation}
\begin{split}
\text{Minimizing } S_T(u)  &= {\int_0}^{T} {\int_{-\infty}}^\infty (u_t-u_{xx}-u^2)^2 dx dt
\label{eq:StartAnalysis}
\end{split}
\end{equation}
subject to the constraints $u(x,0)=0$  and  $u(0,T)=U_f$,
we find that the minimizer $u(x,t)$ satisfies
\begin{equation}
 \begin{split}
 \label{eqn:backForwardBurst}
  u_t= u_{xx} + u^2 + \epsilon \xi, \\
  u(x,0)=0, \\
  -\xi_t= \xi_{xx} + 2 u \xi, \\
  \xi(x,T)= \alpha \delta(x).
\end{split}
\end{equation}

\begin{figure}[h]
    \centering
\includegraphics[width=0.575\columnwidth]{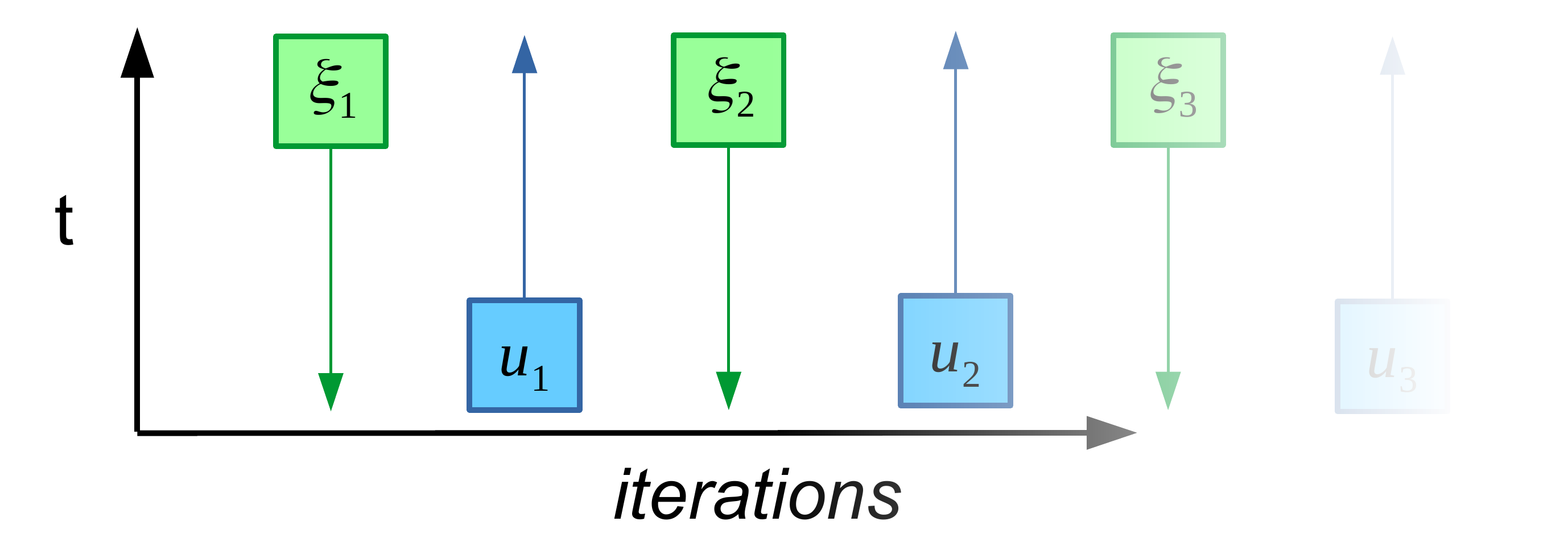}
\caption{Schematic view of our iteration scheme, as used to solve the system of equations (\ref{eqn:backForwardBurst}). }
\label{fig:IterationScheme}
    \end{figure}

Unlike the linear case, where $\xi$ could be solved independently of $u$, and then used to determine $u$ explicitly, here no such analytic solution is available. $\xi$ and $u$ are inextricably coupled.
Solving \ref{eqn:backForwardBurst} numerically in either direction in time requires that we solve the ill posed backwards heat equation, either for $u$ or $\xi$, depending on which direction we solve in. Creating a full spatial-temporal mesh and solving for both $\xi$ and $u$ simultaneously is possible, but quickly becomes computationally expensive for finer meshes.
We can, however, solve iteratively. This is done by initially assuming $u_0(x,t)=0$, and solving $$ -\hat \xi_t = \hat \xi_{xx} + 2u_0 \hat \xi$$ with the boundary condition
 $\hat \xi(x,T)= \delta(x)$, to determine $\hat \xi_1$. We use $\alpha \hat \xi_1$ as a forcing term to find $u_{1}$, and $u_1$ to determine $\hat \xi_2$ and so on, until $|S(u_i)-S(u_{i+1})|<0.0005 |S(u_i)+S(u_{i+1})|$. (See figure \ref{fig:IterationScheme}). 
Using this iteration scheme, we are able to determine $u,\xi$ and $S_T(u)$ for the action minimizing $u$ profile. 
The value of $\alpha$ in each iteration step is found using a bisection method in which $\alpha$ is reduced if $u(0,T)>U_f$ and increased if $u(0,T)<U_f$.

 
 \begin{figure}[h]
    \centering
\includegraphics[width=0.575\columnwidth]{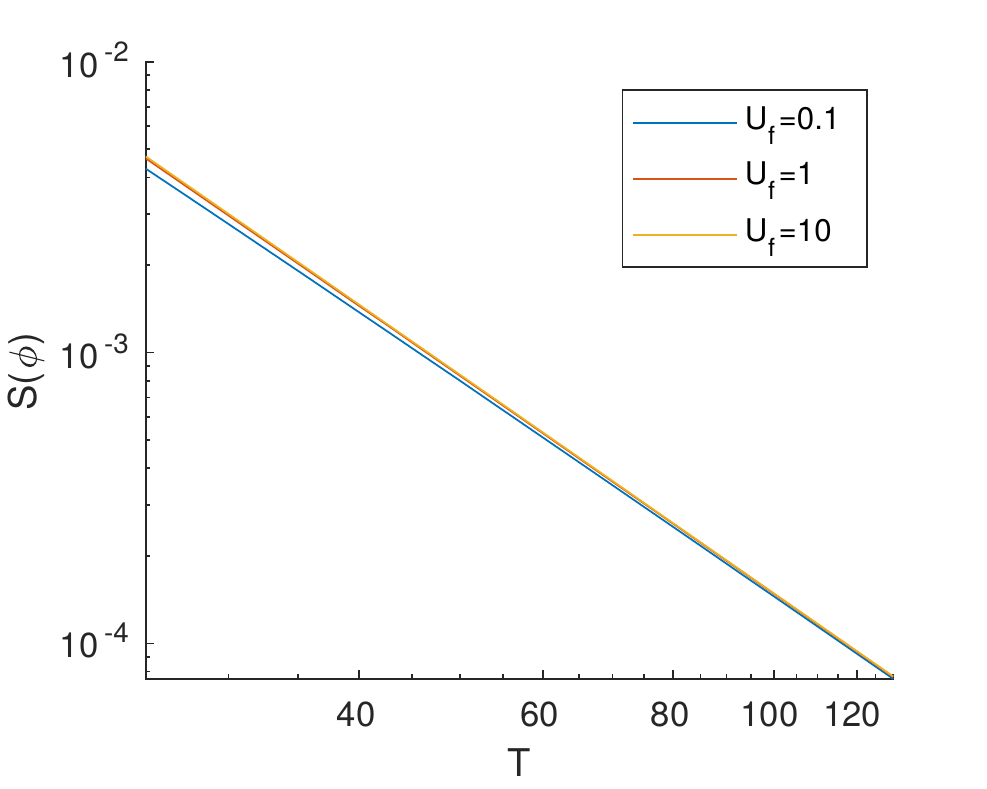}
\caption{Numerical calculation of the action cost $S_T(\phi)$ to reach $u(0,T)=U_f$ by time $t=T$ for equation (\ref{eq:burstWithNoise}), the spatial non-linear case. We consider $U_f= 0.1, 1 $ and $10$; the resulting differences are negligible.
The resulting lines have mean slope $-2.465$, $-2.495$ $-2.498$ in log-log space, which we take to be a good approximation of $5/2$. Note that this $5/2$ result is explicitly an approximation based on numeric results, and it is possible (though presumably not likely) that the exact asymptotic result is some other value.}
\label{fig:ActionCostSpacyModel}
    \end{figure}

\section{Comparisons of results}
\label{sect:compResult}
\subsection{Comparison Between Results for Linear and Non-Linear Case}
\label{sect:compSpaceResult}
Now that we have determined the action minimizing profile in both the linear and non-linear case (equations (\ref{eq:linearBurstWithNoise}) and (\ref{eq:burstWithNoise}), respectively), let us now compare these two models. 

To begin, we would like to determine the time till burst formation in each model. For the sake of notation, let us define $m(T)= \min_\phi S_T(\phi)$. While under normal circumstances we would find the expected escape time by invoking equation \ref{eq:LDT2}, in all cases studied here, $m(T) \rightarrow 0$ as $T \rightarrow \infty$. Hence equation \ref{eq:LDT2} provides us with no information, and we instead rely on equation \ref{eq:LDTmain}.

Regardless of the particular system under study, when $m(T) \gg \epsilon^2$ equation \ref{eq:LDTmain}  implies $-\ln P(\tau \le T) \gg 1$ and hence $P(\tau \le T) \approx 0$. This in turn implies that the probability \emph{density} of burst formation for such values of $T$ is low. Similarly, when $m(T) \ll \epsilon^2$ we have $-\ln P(\tau \le T) \ll 1$ and hence $P(\tau \le T) \approx 1$, and once again the probability of escape is negligible, as escape has almost certainly already occurred.
Escape can only have non-negligible probability density when $\ln P(\tau \le T)$ is neither too big nor too small, namely when 
$m(T)$ and $\epsilon^2$ are  similar orders of magnitude.

This implies that for the linear model escape is predicted when $\epsilon^{2} \approx \bar S_T = U_f^2 O(e^{-2T})$, or equivalently when $T \sim -\log(\epsilon)$. The time of escape changes little, even as $\epsilon$ is varied over several orders of magnitude.
In the nonlinear case we observe that $m(T)= O(T^{-5/2})$ as $T \rightarrow \infty$ (see figure \ref{fig:ActionCostSpacyModel}).
By the above arguments, we predict $T \sim \epsilon^{-4/5}$; the time till burst formation is \emph{sensitive} to the amplitude of the noise driving the system. 
Table \ref{table:ta} summarises these timing results, and compares to the previously discussed spaceless models.

\begin{figure}
    \centering
\includegraphics[width=0.475\columnwidth]{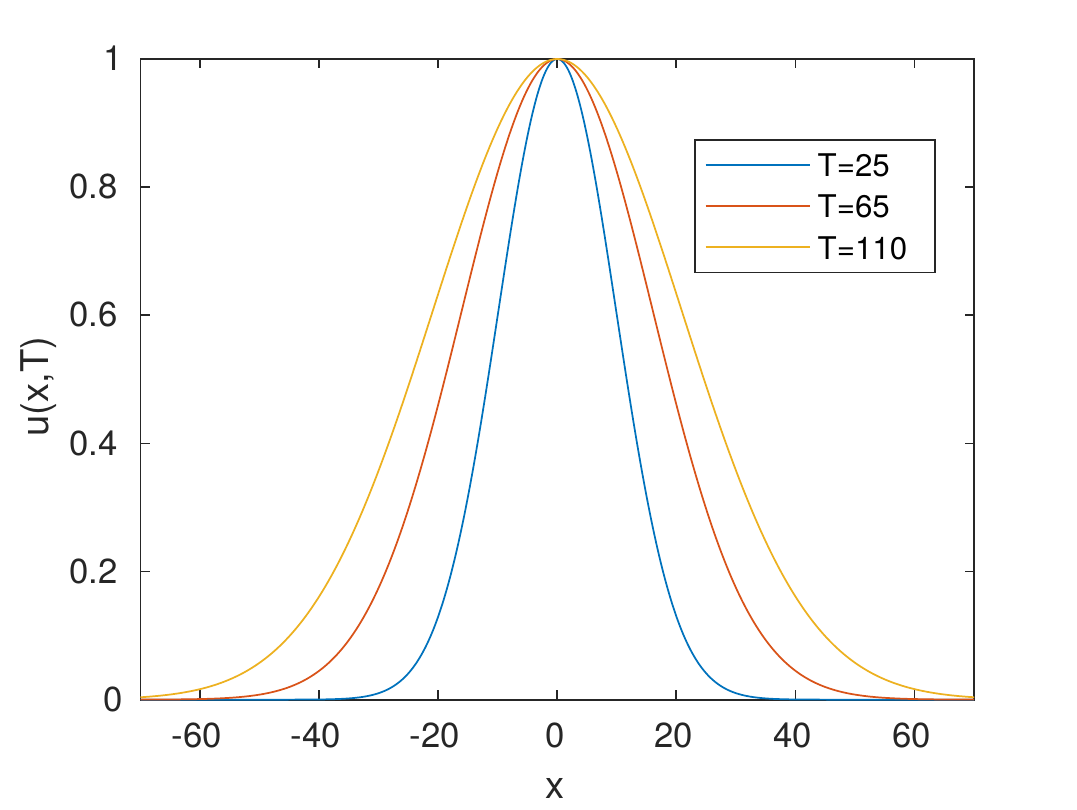}
\includegraphics[width=0.475\columnwidth]{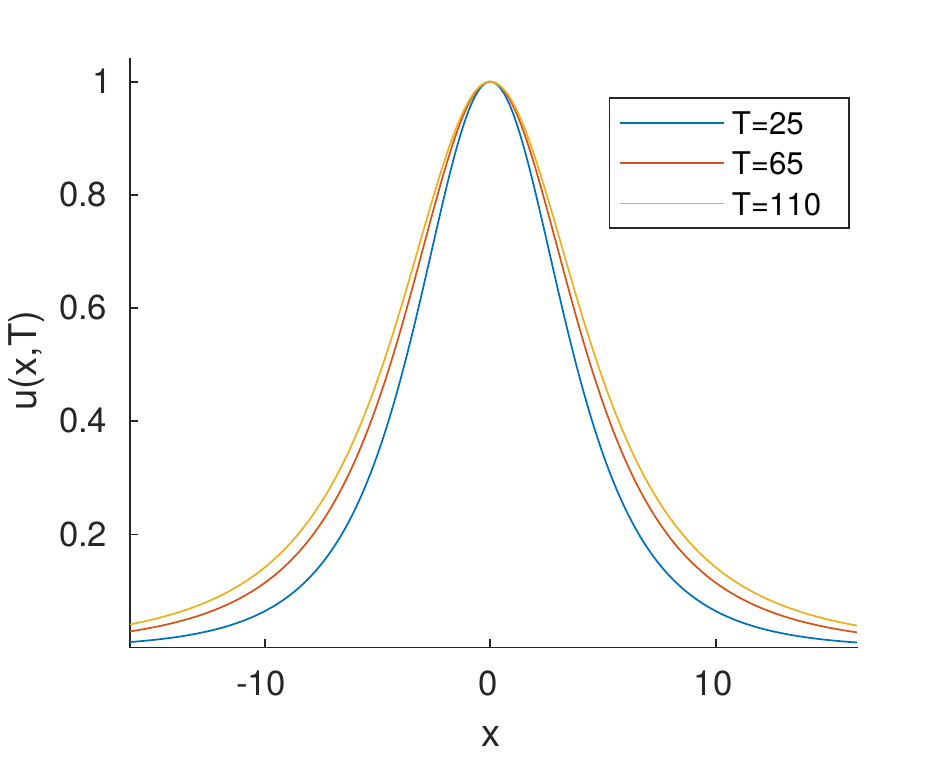}
\label{fig:FinalProfiles}
\caption{(Left) In the case of linear instability (equation \ref{eq:linearBurstWithNoise}) The profile of $u(x,T)$ is well approximated by $N(x,2T)$. The width of our peak varies significantly depending on $T$ . (Right) In the case of quadratic instability (equation \ref{eq:burstWithNoise}), as might be observed in the vicinity of a saddle-node bifurcation, the profile of $u(x,T)$ is relatively insensitive to $T$, and appears to approach some limiting ``canonical burst'' as $T$ is made large.}
    \end{figure}

Far more interesting than the difference in escape times for the linear and quadratic case is the substantial difference in the profile of our action-minimizing $u$. In the linear case, $u(x,t)$ was found explicitly (equation \ref{eq:bestLinearBurst}). In this case $u(x,T)$ is well approximated by a corresponding normal distribution $N(x,2T)$, and becomes wider over time (see figure \ref{fig:FinalProfiles}, upper). In contrast, in the quadratic case, the final width of $u$ increases at most very slowly, and appears to approach a limiting distribution as $T \rightarrow \infty$ (see figure \ref{fig:FinalProfiles}, lower). 
 
\begin{figure}[h]
    \centering
\includegraphics[width=0.475\columnwidth]{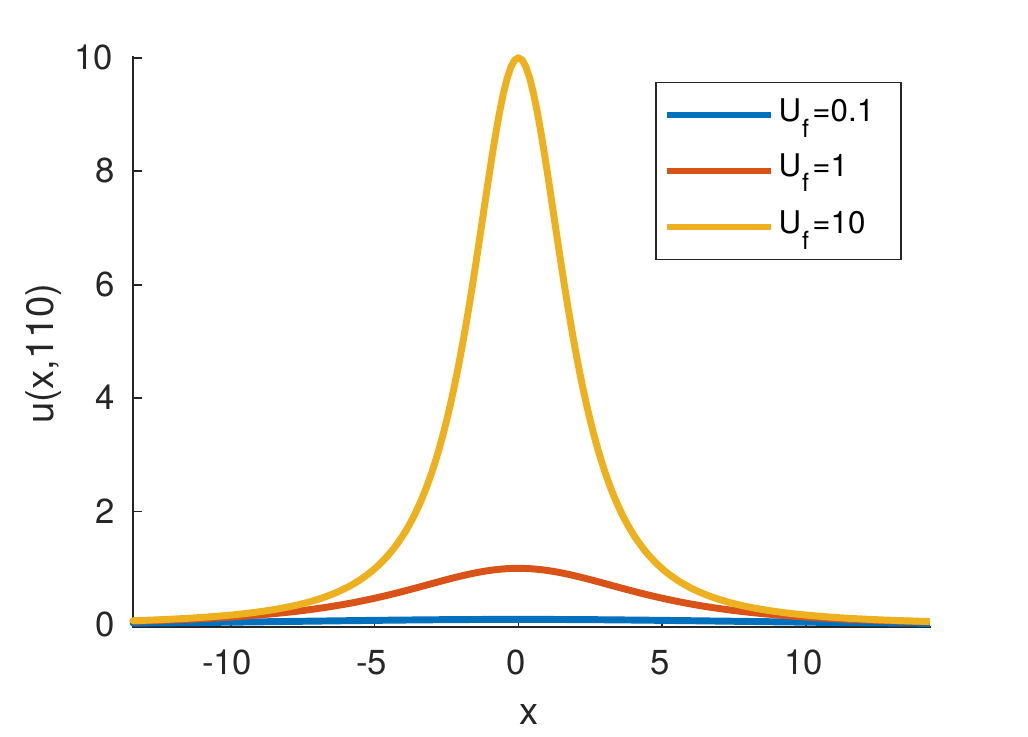}
\includegraphics[width=0.475\columnwidth]{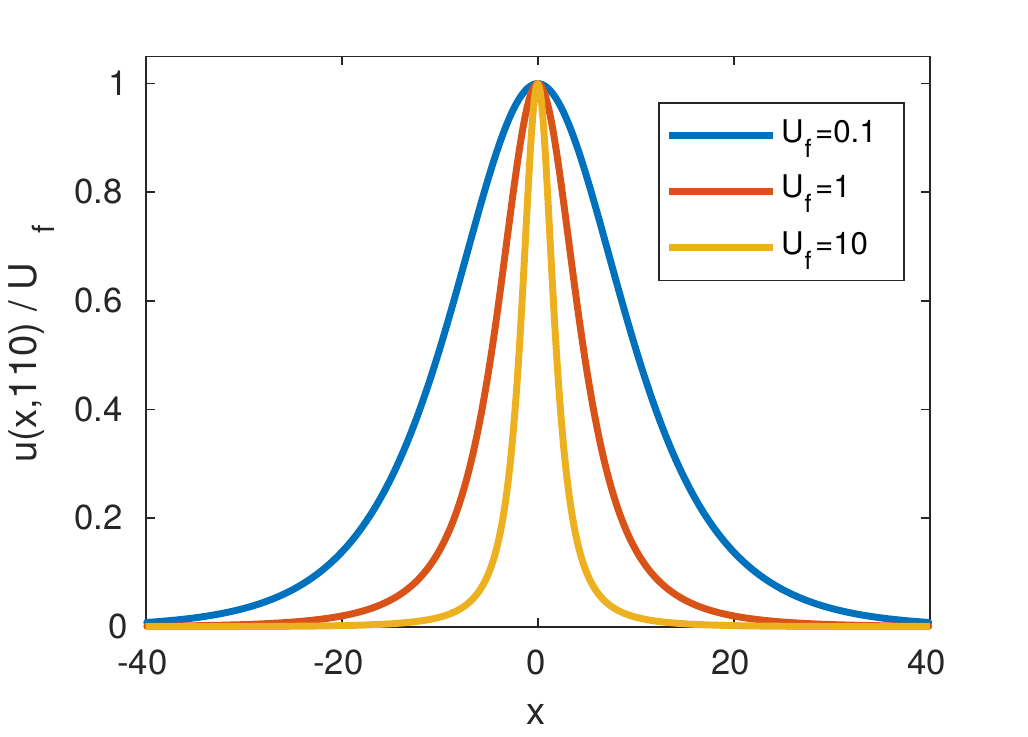}
\caption{(Left) For fixed $T=110$, we find $u(x,t)$ for a variety of $U_f$. As $U_f$ is increased, $u(x,T)$ increases monotonically, however the most significant gains happen near $x=0$, and hence, the profile becomes more `spike' like, as can be seen in the amplitude normalized plots (Right) }
\label{fig:SpikeProfileVaryU}
    \end{figure}

As a final point of comparison, we consider the reaction of the two systems to changes in $U_f$. As might be expected, the \emph{shape} of the action-minimising $u$ profile is unaffected by changes in $U_f$ for the linear system. Doubling $U_f$ for a given $T$ doubles both $\xi$ and $u$, but does not change the shape of either.
In contrast, in the non-linear case, the profile of $u(x,t)$ is sensitive to changes in $U_f$, with larger $U_f$ leading to a sharper ``spike'' solution, and smaller $U_f$ leading to soft ``bumps'' (see figure \ref{fig:SpikeProfileVaryU}).

\begin{table}[ht]
\centering
\begin{tabular}{ c | c | c  }

Equation & $m(T)$ & $\mathbb{E}(T) \sim$ \\
\hline
$u_t = u^2 + \epsilon\xi$ & $O(T^{-3})$ & $\epsilon^{-2/3}$ \\
$u_t = u^2 + \gamma^2 + \epsilon\xi$ & 0 & $\gamma^{-1}$ \\
$u_t = u^2 - \gamma^2 + \epsilon\xi$ & $16\gamma^3/3$ & $\exp[\epsilon^{-2} 16\gamma^3/3]$ \\
$u_t = \Delta u + u + \epsilon\xi$ & $U_f O(e^{-2T})$ & $\log(\epsilon) $ \\
$u_t = \Delta u + u^2 + \epsilon\xi$  & $O(T^{-5/2})$ & $\epsilon^{-4/5}$ \\
\end{tabular}
\caption{Comparison of action functionals and approximate escape times.}
\label{table:ta}
\end{table}

\subsection{Comparison of Analytic Results to Direct Simulations}
As a final check on the above analytic results, we compare to simulations of the corresponding systems. 

\begin{figure}[h]
    \centering
\includegraphics[width=0.475\columnwidth]{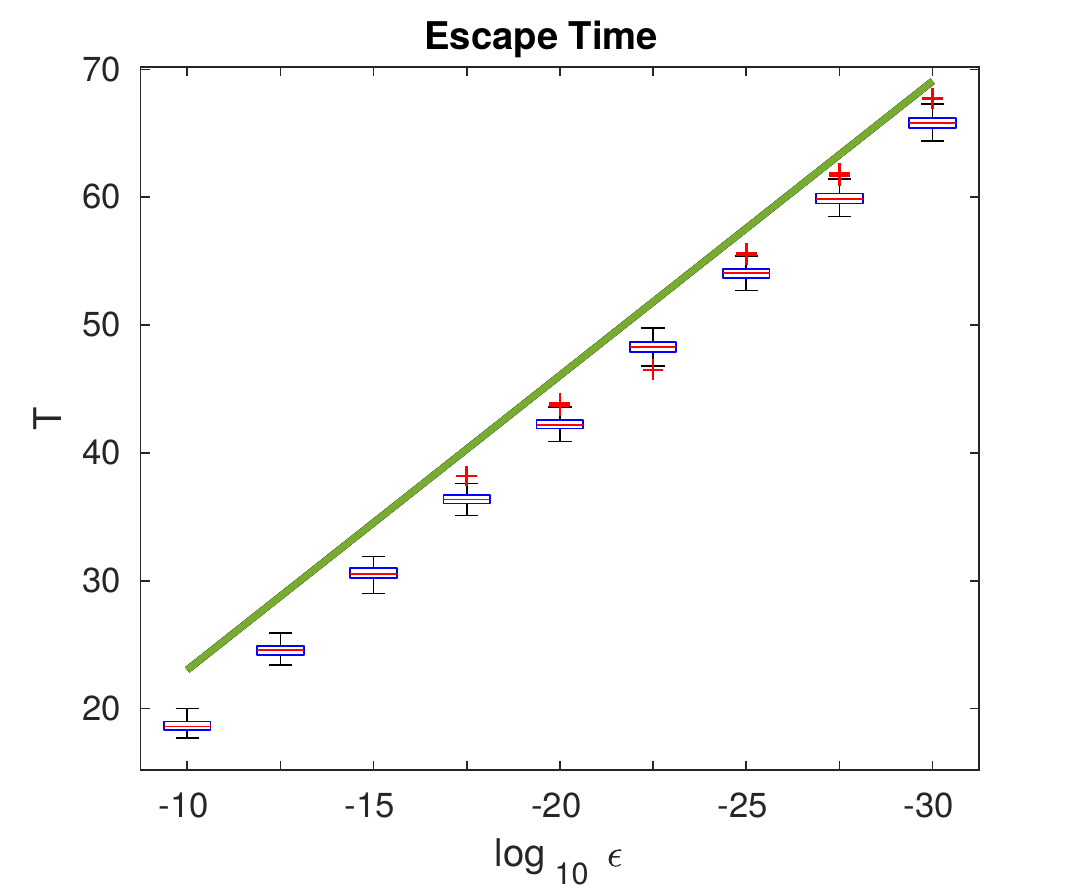}
\includegraphics[width=0.475\columnwidth]{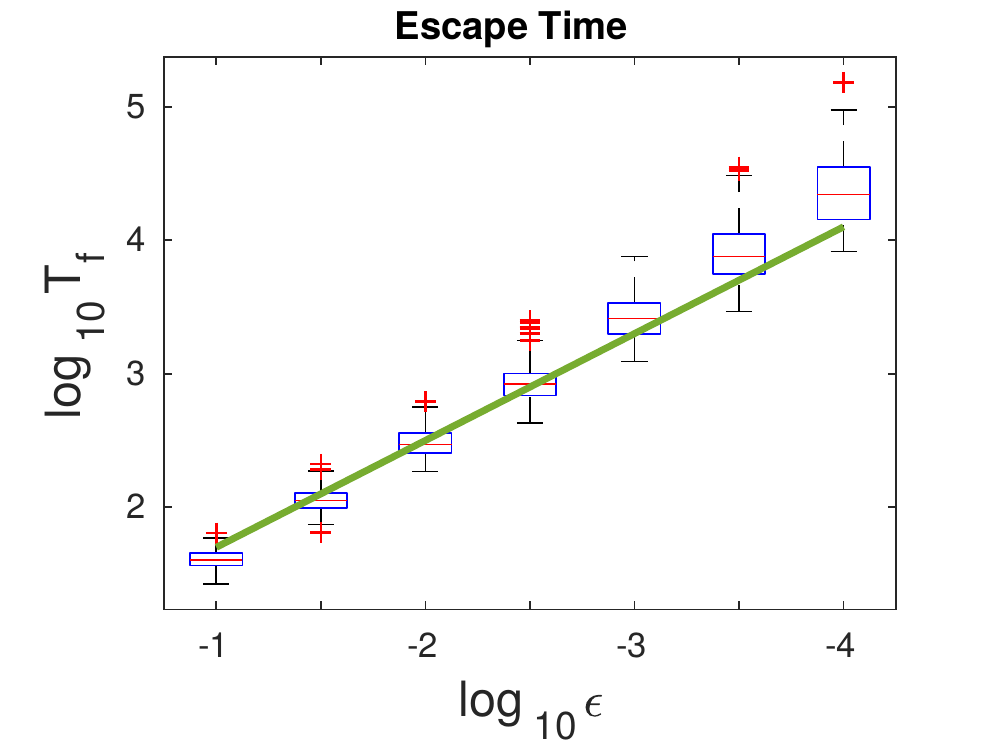}
\caption{(Left) Passage times from $u(x,0)=0$ to $\max|u(x,t)|=1$, for equation \ref{eq:linearBurstWithNoise}. For each $\epsilon$ value we use $300$ simulations. Box plots indicate simulated passage times, while the solid green line indicates the predicted scaling $T \approx -log(\epsilon)$. Given that theory only guarantees this approximation correct up to an order of magnitude, the observed agreement between theory and simulation is gratifying. The vertical shift between theory and simulation is a reflection of the unknown coefficient in the scaling relationship. (Right) Similar simulations for equation \ref{eq:burstWithNoise}. For each $\epsilon$ value we use $250$ simulations. Box plots indicate simulated passage times, while the solid green line indicates the predicted scaling $\log(T) \approx -4/5 \log(\epsilon)$. While predictions are reasonable, linear regression indicates that the best fitting slope is $-0.924344 \pm 0.003856$, however even this assertion must be treated with some skepticism- as the graph does not appear to obey a linear relationship for small values of $\epsilon.$
Given that the $\log(T) \approx -4/5 \log(\epsilon)$ prediction is based on extrapolation from numerical results, it is unclear whether the discrepancy comes from inaccuracies in our simulation method or is a result of genuine limitations on the $-4/5$ relation itself.}
\label{fig:simLinearEscape}
    \end{figure}

In the linear case (equation \ref{eq:linearBurstWithNoise}), we find that the time until $\max |u(x,t)| = U_f = 1$ scales like $-\log(\epsilon)$, as predicted by theory (see figure \ref{fig:simLinearEscape}, left).    
In the non-linear case (equation \ref{eq:burstWithNoise}), burst time is reasonably approximated by $\epsilon^{-4/5}$ as predicted by theory, but the agreement is weaker than might be hoped (see figure \ref{fig:simLinearEscape}, right). It is unclear if this disagreement stems from the limitations of our simulations, or the numeric-analytic arguments proposed earlier.

\begin{figure}[h]
    \centering
\includegraphics[width=0.475\columnwidth]{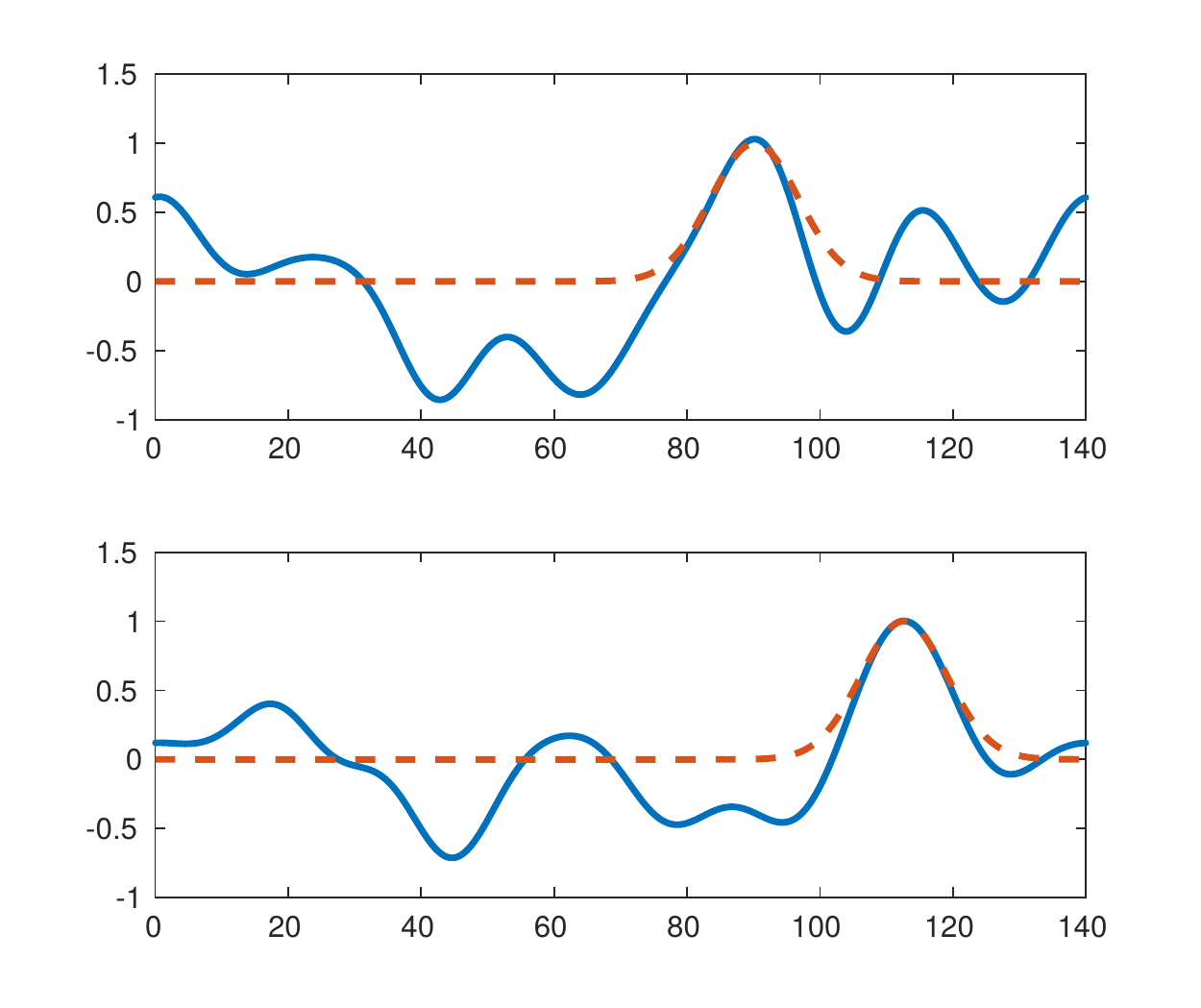}
\includegraphics[width=0.475\columnwidth]{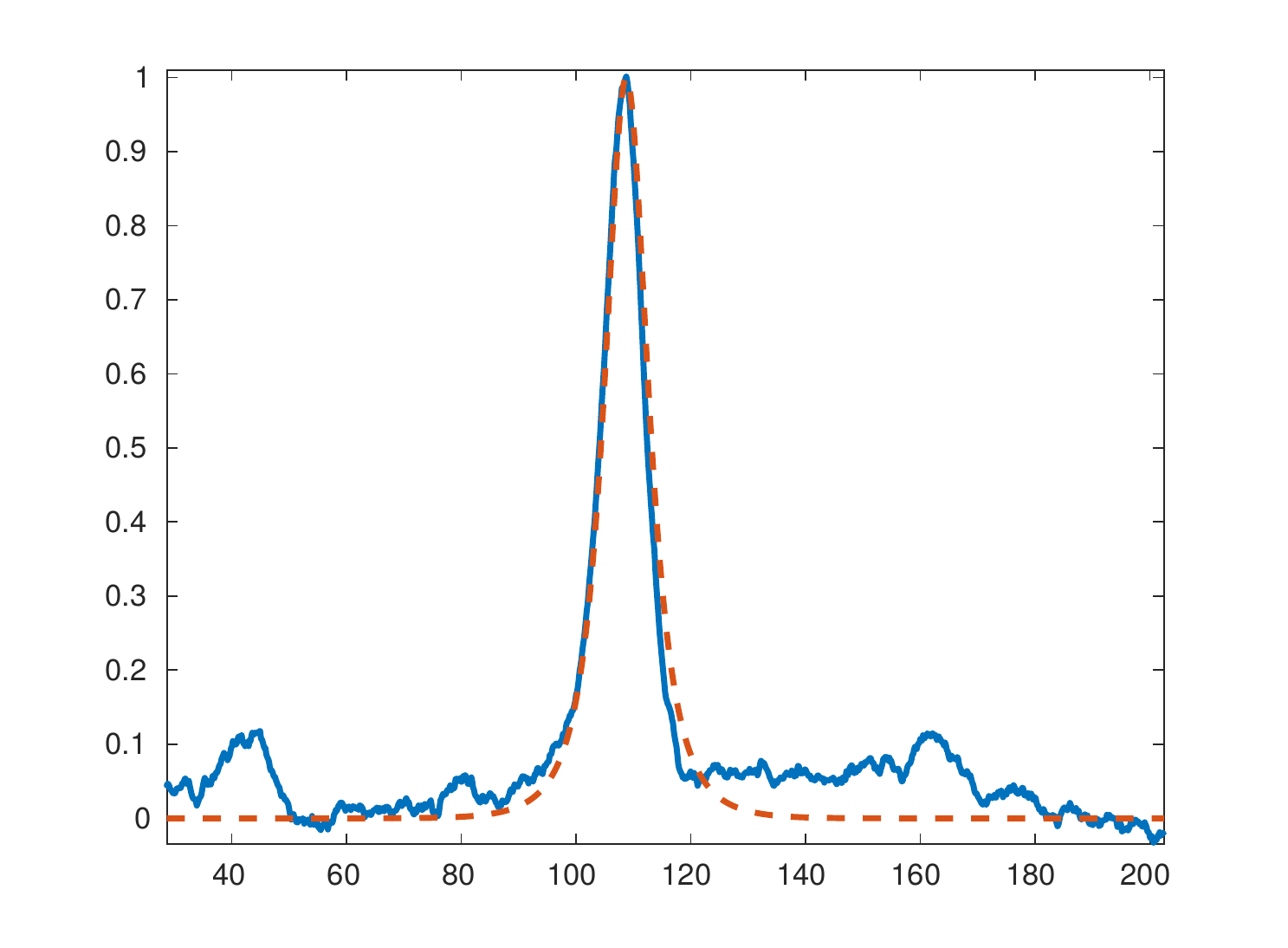}
\caption{(Left) Final profiles of $u(x,T)$ (solid line), along with the accompanying ``ideal burst'' (dashed line) for two simulations of equation \ref{eq:linearBurstWithNoise}.
(Right) Final profile of $u(x,T)$ (solid line), along with the accompanying ``ideal burst'' for a single simulation of equation (\ref{eq:burstWithNoise}) with $\epsilon=0.1$.
In both cases, no fitting was used beyond a horizontal shift moving the peak of our analytic solution in line with the peak realized in simulation. Peak shape is determined entirely by the asymptotic arguments previously described.}
\label{fig:simLinearShape}
    \end{figure}

Comparison of shape in the linear case finds reasonable agreement in the vicinity of the primary burst itself, however because a large number of bursts form simultaneously, the final profile of $u$ ends up being an overlapping combination of many different such bursts (see figure \ref{fig:simLinearShape},left).

In contrast, comparison between the predicted and observed shape of bursts shows strong agreement in the non-linear case, even for relatively large values of $\epsilon$ (see figure \ref{fig:simLinearShape},right). Bursts spend a greater portion of their development time close to zero in the non-linear case, and hence, although many bursts may begin growing at roughly the same time, the second and third place bursts will have only a minor effect on the overall profile of $u(x,T)$. 

The code for all simulations can be found in the supplementary materials.

\section{Difficulties in 2D, and higher dimensions}
\label{sect:twoDbad}
Given that the universe we occupy is not one dimensional, it would be beneficial to extend the above results to higher dimensions, preferably three dimensions, although for certain contexts, such as the binding of proteins to a cell membrane\cite{vecchiarelli_membrane-bound_2016}, two dimensions would suffice.

Unfortunately, the properties of white noise preclude this avenue of investigation; as noted by Ryser et al \cite{ryser_well-posedness_2012}, equations of the form $u_t=\Delta u + \xi$ do not remain well posed in dimension two or higher. The smoothing effects of diffusion are insufficient to restrain the irregularity produced by space-time white noise.

This result can be seen through a variety of lenses, depending on ones outlook. Ryser et al, describe $u$ in terms of a sum of Fourier modes, and show $$\mathbb{E}||u||^2 = \sum \epsilon (1- e^{-2(1+|{\bf k}|^2)t})/2 (1+|{\bf k}|^2).$$ In one dimension, when ${\bf k} \in \mathbb{Z}$ this sum converges, however for two or more dimensions, ${\bf k} \in \mathbb{Z}^d$ and the sum is unbounded for all $t>0$. This indicates that $u$ is not well defined.
Simulations in 2d, such as those presented in section \ref{sect:compSpaceResult} function for fixed $dx$, but invariably break down as $dx \rightarrow 0$.

In terms of the methodology used in this paper, the degeneracy of higher dimensions manifests as a break down of various integrals. Issues arise regardless of what equation is used, but the problem is most easily illustrated by considering the linear case.
Consider the solution $u(x,t)$ of eq. \ref{eq:linearBurstWithNoise} given in eq. \ref{eq:bestLinearBurst}. The parameter $\alpha$ of eq. \ref{eq:bestLinearBurst} is determined by considering the boundary condition $u(0,T)=U_f$, 
$$U_f =\frac{ \alpha \epsilon}{2} \int_0^{2T} e^\gamma N(0,\gamma) d\gamma,$$ where $\gamma=T+t-2\tau$.

In 1D $ N({\bf x},\gamma)=  (4 \pi \gamma)^{-1/2} \exp \left[-x^2/4\gamma \right]$, and hence $U_f = \frac{\alpha \epsilon}{4\sqrt{\pi}} \int  e^\gamma \gamma^{-1/2} e^{-0/\gamma} d\gamma$. Because the integral converges to a finite number, we can rearrange to determine $\alpha$ and go on to find $S(u)$ explicitly for any $U_f$. 

By contrast, in 2D we have $ N({\bf x},\gamma)=  (4 \pi \gamma)^{-1} \exp \left[-\|{\bf x}\|^2/4\gamma \right]$, and hence, $U_f = O(\alpha \epsilon) \int e^\gamma \gamma^{-1} e^{-0/\gamma} dx$. The integral is unbounded, and thus we are able to select $\alpha$ arbitrarily small. This in turn allows both $T$ and $S(u)$ to be selected arbitrarily close to zero, regardless of $U_f$. Physically, we are picking $\xi$ to be a spike that is sufficiently narrow so as to ensure $S(u)$ arbitrarily small, yet also sufficiently high so as to force $u$ from zero to $U_f$ in time $T$. This sleight of hand is impossible in one dimensional systems, but for all higher dimensions it is the ``action minimizing'' strategy.

While here we have described the difficulties of integration in the linear case, the problems described extends to all equations that include noise terms in 2D, and become progressively worse for higher dimensions. They are a symptom, rather than the root cause of the problems of noise in higher dimensions.

\section{Conclusions, Discussion and Future work}
\label{sect:Conclude}
In this article we have presented a number of prototypical models exploring the behavior of systems starting at unstable and semi-stable steady states.

We have demonstrated both analytically and through simulations that under the influence of spatio-temporal white noise, systems of the form $u_t = u_{xx} +u^2 +\epsilon \xi$ develop well defined ``spikes'' in finite time. This phenomena can be seen as delayed bifurcation \cite{kuehn_mathematical_2011} in a spatially distributed system, and also as a compliment to non-linear blow up phenomena, in which we consider instead the system's ``escape from zero'', as opposed to the more typically studied ``approach to infinity'' \cite{vazquez_problem_2002}. 

The symmetry breaking observed here does not rely on multiple chemical species, nor on contrasts in diffusion rate, as might be expected for the more classical Turing type pattern formation.

We were able to determine both the ``energy cost'' of spike formation, along with the shape of the most probable spike, and showed that in the limit of large time, spike formation requires infinitesimally small energy. The sensitivity of the system in the vicinity of its semi-stable steady state allows noise of amplitude $\epsilon \ll 1$ to have macroscopic effects in O($\epsilon^{-4/5}$) time.
We also explored similar results both in the spaceless case, and in the case where our $u=0$ state is linearly unstable as opposed to semi-stable. 

In terms of physical relevance, the systems studied can be best thought of as a prototype for chemical reaction-diffusion systems in the vicinity of a saddle-node bifurcation. Our study demonstrates the breakdown of a homogeneous state, and can be seen as a first step towards understanding systems such as the Min system studied by Vecchiarelli et al \cite{vecchiarelli_membrane-bound_2016}, in which changes in the bulk concentration of a protein pushes the system through a bifurcation boundary and leads to the formation of membrane bound ``bursts''.

There are a number of further questions which must be answered before the work here can be applied to any experimental context. First and foremost, the work here considers a system perfectly balanced at the saddle-node; $u_t=u^2$. No physical system however is ever so perfectly balanced, and so determining the robustness of these results for a system that is \emph{passing through} such a saddle node bifurcation is critical to our understanding of the relevance of these results. Further in order for the results here to prove useful for experimental scientists, time must be invested in investigating the mapping between experimental observations, and the associated reaction and diffusion parameters implied; for example, does the relationship between spike width and height give a reliable signature that can be used to determine system parameters? 

Finally, our work here alludes (in passing) to potentially deeper questions in chemistry; namely the difficulties in standard representations of noise when in higher dimensions. Such questions must be answered, or at the very least sidestepped, before the phenomena observed here can be sensibly applied to the multi dimensional systems ubiquitous in the real world.

\section{Code}
The Code used in this project is available on github at alastair-JL/StochasticBurst.

\section{Acknowledgements}
We wish to acknowledge Anthony Vecchiarelli, for stimulating discussion on the topic of Min proteins, and sharing his data. This research was funded by the four year fellowship from the university of British Columbia, and NSERC.

\appendix
\section{Introduction to Large Deviation Theory}
\label{ap:IntroToLDT}
Here we give a very brief introduction to the principle ideas and techniques used in Large Deviation Theory (LDT), and used throughout this paper.
In this appendix we present LDT as it applies to discrete time Stochastic processes. The continuous time case is conceptually similarly. Our goal here is to build intuition rather than mathematical rigor.  

As a concrete example with which to frame our discussion, consider the Discrete Orenstien-Ulembeck process\cite{larralde_first_2004}
\begin{align}
u_i= 0.9 u_{i-1} + \epsilon \xi_i,\\
u_0=0.
\label{eq:OU}
\end{align}
Here $\xi_i$ is assumed to be a normal random variable with mean zero and variance one, such that each $\xi_i$ is independent. We assume $0<\epsilon \ll 1$. 

Suppose we wish to know the first time that $u_i>1$. At first glance, given $O(\epsilon)$ noise, and the decay term of our OU process, $u_i>1$ seems unlikely to occur. That said, as we take $i \rightarrow \infty$, unlikely, even exceedingly unlikely events should occur \emph{eventually}.
LDT concerns itself with such questions as ``how long will it take for X to occur?'' and ``given that X occurs, what path is the system most likely to take in order to get there?''. In our particular case X is  ``$u_i>1$''.

The total probability of our event $X$ is equal to the integral over the probability density of all paths leading to that event. This integral can be formulated either in terms of the paths of the stochastic process ${\bf u}$, or in terms of the underlying noise $\boldsymbol \xi$.
\begin{align}
P(X)=\int_{u \in X} p_u({\bf u}) d{\bf u} = \int_{\xi \in X} p_\xi({\boldsymbol \xi}) d\bf {\boldsymbol \xi}.
\end{align}
The probability density function for a noise vector $\boldsymbol \xi$ of length $N$ is
\begin{align}
p_\xi({\boldsymbol \xi})= (2\pi)^{-N/2} e^{-\sum \xi_i^2 /2}.
\label{eq:xiVectorDensity}
\end{align}

Unfortunately, even with this well defined probability density, the boundary of the integral $\int_{\xi \in X} p_\xi({\boldsymbol \xi}) d\bf {\boldsymbol \xi}$ is generically complicated enough so as to prevent us from evaluating $P(X)$ directly (the OU process being a notable exception).

In order to avoid this complex integration boundary, it is useful to transform our integral back into a form dependent on ${\bf u}$.  To achieve this we rearrange our recurrence relation (equation \ref{eq:OU}), and find
\begin{align}
\xi_i = \frac{u_i-0.9u_{i-1}}{\epsilon}.
\end{align}
Using this one-to-one correspondence between ${\bf u}$ and ${\boldsymbol \xi}$ along with the change of random variables formula, we find
\begin{align}
P(X) &=\int_{u \in X} p_u({\bf u}) d{\bf u} = \int_{\xi \in X} p_\xi({\boldsymbol \xi}) d\bf {\boldsymbol \xi}\\
&=(2\pi)^{-N/2} \int_{\xi \in X} \exp \left[-\sum \xi_i^2 /2 \right] d\boldsymbol \xi\\
&=(2\pi)^{-N/2} \int_{u \in X} \exp \left[-\sum (u_i-0.9u_{i-1})^2 /2\epsilon^2 \right] d\bf u\\
&=(2\pi)^{-N/2} \int_{u \in X} e^{-S({\bf u})/\epsilon^2} d\bf u.
\end{align}
Here $S({\bf u})$ is said to be the ``normalized action functional'' of our problem. In the particular case discussed here $S({\bf u})= \sum (u_i-0.9u_{i-1})^2 /2$. In general $S({\bf u})$ depends both on the equation governing a stochastic process, and the particular form of the noise generating it.  
$S({\bf u})$ can be thought of as a measure of the ``total improbability'' associated with a given path, and is associated with the amount of ``energy'' that noise must pour into the system in order to cause a particular path to occur. When dealing with continuous systems, $S$ is defined as an integral over the square of noise, rather than a sum.

At this stage, in order to determine the probability of our event, we need to evaluate $\int_{u \in X} e^{-S({\bf u})/\epsilon^2} d {\bf u}$. Typically, this integral can not be evaluated exactly, but it can be well approximated via Laplace's Principle \cite{laplace_memoir_1986,olivieri_large_2005}.

\begin{figure}[h]
    \centering
\includegraphics[width=0.475\columnwidth]{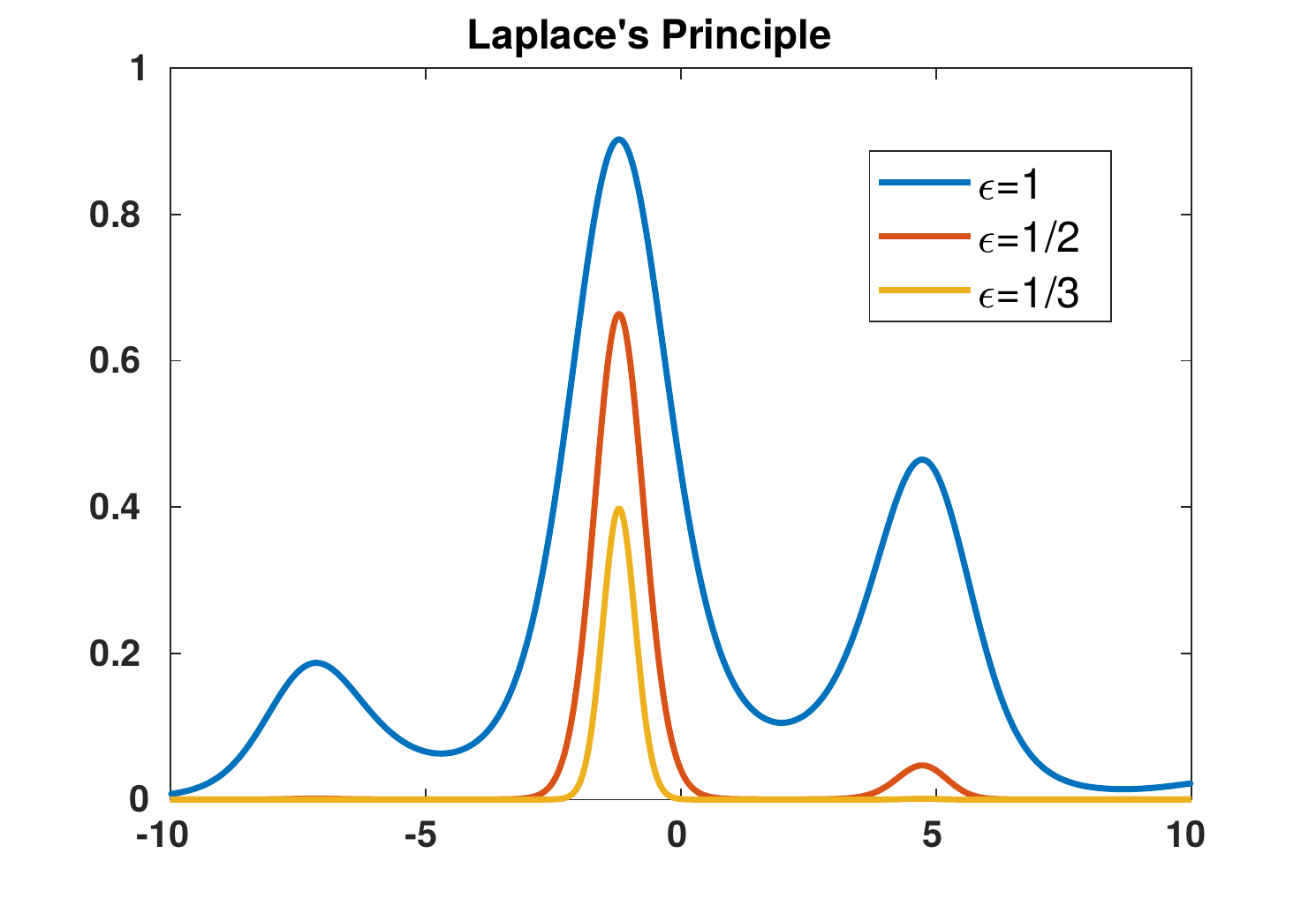}\\
\includegraphics[width=0.475\columnwidth]{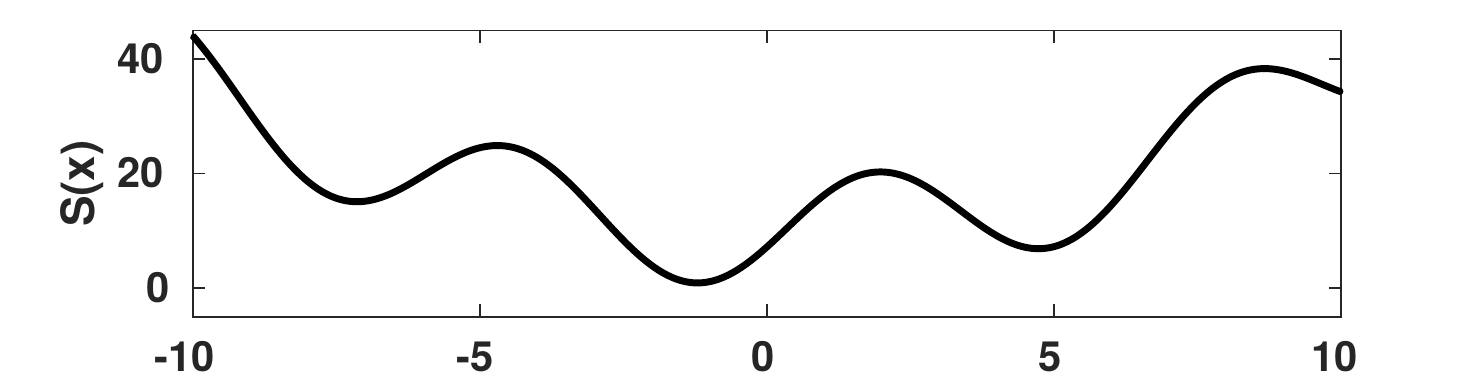}
\caption[Simple Illustration of Laplace's Principle]{For some arbitrary function $S(x)$ , as $\epsilon\rightarrow 0$ the vast majority of the integral $\int e^{-S(x)/\epsilon^2} dx$ can be found in a narrow window near the minimum of $S(x)$. 
As a result $\int e^{-S(x)/\epsilon^2} dx$ scales like $e^{-S(x_0)/\epsilon^2}$. In this particular example, for $\epsilon=1$ probability mass is spread across a number of separate ``peaks'' (local minima of $S(x)$). For $\epsilon=1/3$ by contrast, only the global minima of $S(x)$ has non-negligible mass associated with it. While varying $\epsilon$ does change the width of our global peak, this effect is negligible compared to the change in height resulting from $e^{-S(x_0)/\epsilon^2}$.
}
\label{fig:LaplacePrinceCartoon}
    \end{figure}

Laplace's Principle states that:
\begin{align}
\int_{u \in X} e^{-S({\bf u})/\epsilon^2} d {\bf u} \sim \exp \left[ -\min S({\bf u})/\epsilon^2 \right],
\end{align}
where here we minimize $S({\bf u})$ over all ${\bf u} \in X$. 
Laplace's principle is based on the idea that for integrals of the from $\int_{u \in X} e^{-S({\bf u})/\epsilon^2} d {\bf u}$, the overwhelming majority of all probability mass is concentrated in the vicinity of $\min S({\bf u})$ whenever $\epsilon \ll 1$, (see figure \ref{fig:LaplacePrinceCartoon}). From the point of view of stochastic processes, what Laplace's Principle is effectively stating is that if an improbable event does occur, the observed path is overwhelmingly likely to be `close' to the most probable path.

Let us return to the original question proposed at the start of this section: how long does it take before the Discrete Orenstien-Ulembeck process defined in equation \ref{eq:OU} exceeds one for the first time?

Suppose we wish to determine the probability that $u_N\ge 1$, for some particular $N$, assumed to be large. By Laplace's Principle, we must thus minimize $S({\bf u})= \sum (u_i-0.9u_{i-1})^2 /2$. Because any $u_N > 1$ will only increase $S({\bf u})$ we can assume $u_N=1$. 
In order to minimize, we require that $u_N=1, u_0=0$ and $\frac{d S}{du_i} = 0$ for all $i$.
As $N\rightarrow \infty$, solutions to the above can be well approximated by $u_i= 0.9^{N-i}$. Substituting back into the definition gives $S({\bf u})=0.095$, and hence $P(u_N\ge1)\approx \exp[-0.095/\epsilon^2]$. If this is the probability of success for any particular large $N$, then we infer that the expected time until $u_i>1$ will scale such that
$\mathbb{E}(\tau)= O(e^{0.095/\epsilon^2})$.

Freidln and Wentzell\cite{freidlin_random_2012}  provide the general formulation for the above two results.
They state (chapter 4, theorem 1.2) that for small epsilon the probability of a particular rare event occurring by a given time $P(\tau<T)$ is governed by:
\begin{equation}
\lim_{\epsilon \rightarrow 0} \epsilon^2 \ln P(\tau \le T) = - \min_\phi S_T(\phi).
\label{eq:LDTmainDuplicate}
\end{equation}
Here, as is customary, $\phi$ denotes a \emph{particular} trajectory of $u$ and $S_T(\phi)$ indicates that we are minimising over all trajectories $\phi$ such that our rare event occurs by time $T$.
This theorem is referred to as equation (\ref{eq:LDTmain}) in the main text.

\section{Derivation of analytic expression for $\bar S_T$ in the spaceless case.}
\label{app:CunningByParts}
Here we present the derivation of equations \ref{eq:FinalSpacelessAction}.
We begin by taking the definition of $\bar S_T$ and expanding.
From eq. \ref{eq:MinimizationDefine} we can show $\dot \phi = \sqrt{\phi^4 +C}$ and hence:

\begin{align}
\bar S_T(\phi) = & \int_0^T \dot \phi^2 - 2 \dot\phi \phi^2 +\phi^4 dt\\
= & \int_0^T \dot \phi^2 - 2 \dot\phi \phi^2 + \phi^4 +C -C dt\\
= & 2 \int_0^T \dot \phi \sqrt{\phi^4 +C} - \dot\phi \phi^2 dt - \int_0^T C dt\\
= & 2 \int_0^{U_f} \sqrt{\phi^4 +C} d\phi  -  2\int_0^{U_f} \phi^2 d\phi - \int_0^T C dt.
\end{align}
The problematic term here is the first integral. Applying integration by parts gives

\begin{align}
\int_0^{U_f} \sqrt{\phi^4 +C} d\phi = & \left[\phi \sqrt{\phi^4 +C} \right]_0^{U_f} - \int_0^{U_f} \phi \frac{4 \phi^3}{ 2 \sqrt{\phi^4 +C}} d\phi, \\
= & \left[\phi \sqrt{\phi^4 +C} \right]_0^{U_f} - \int_0^{U_f} 2 \frac{ \phi^4+C-C}{\sqrt{\phi^4 +C}} d\phi. \\
\int_0^{U_f} 3 \sqrt{\phi^4 +C} d\phi = & \left[\phi \sqrt{\phi^4 +C} \right]_0^{U_f} + \int_0^{U_f}  \frac{2C}{\sqrt{\phi^4 +C}} d\phi.
\end{align}
By eq. \ref{eq:Cpinning} this last term is $2 C T$. Substituting this result back in gives.
\begin{align}
\bar S_T(\phi) =& \frac{2}{3} \left[\phi \sqrt{\phi^4 +C} \right]_0^{U_f} + \frac{4}{3} C T  -  2\int_0^{U_f} \phi^2 d\phi - \int_0^T C dt,\\
=& \frac{2}{3} \left[\phi^3 \sqrt{1 +C \phi^{-4}} \right]_0^{U_f}  -  \frac{2}{3} \left[\phi^3 \right]_0^{U_f} + \frac{1}{3} C T,\\
=& \frac{2}{3} \left[\phi^3 \left( \sqrt{1 +C \phi^{-4} } -1 \right) \right]_0^{U_f}  + \frac{1}{3} C T,
\end{align}
hence recovering equation \ref{eq:FinalSpacelessAction}.

At no point in this derivation have we used the approximation of $C$ given in eq \ref{eq:Capprox}, hence this is an analytic result.

\section{The linearly unstable case; full calculation}
\label{Ap:AlgebraDetail}
Here we provide the detailed algebra suppressed in section \ref{subsect:LinUnstable}. 
Consider the minimization problem:
\begin{align}
\label{eq:Base_SDE_Appendix}
\text{Given  } &u_t = u_{xx} + u + \epsilon \xi, \nonumber\\
& u(x,0)=0, \nonumber\\
\text{minimise } & S(u(\xi)) = {\int_0}^{T} {\int_{-\infty}}^\infty \epsilon^2 \xi^2 dx dt \nonumber\\
\text{such that  }& u(0,T)=U_f.
\end{align}
This is equivelent to equation (\ref{eq:StartAnalysis}) of the main text.

If we consider $\xi$ as a form of forcing function, then the above asks for the lowest energy forcing required to lift $u(0,T)$ to height $U_f$. Thinking of $\xi$ as a form of noise we are asking for the \textit{most probable} noise.  For the rest of this appendix we will think of $\xi$ in terms of energy input, although we will keep the white noise formulation in mind.
Please note that while both $u$ and $\xi$ are stochastic processes, in a slight abuse of notation we also use the symbols to indicate the optimal path subject to the above constraints.
\\

In order to find our minima, we first re-write $S$ in terms of $u$, and then take the functional derivative:

\begin{align}
  S = {\int_0}^{T} {\int_{-\infty}^\infty}  \epsilon^2 \xi^2 dx dt = {\int_0}^{T} {\int_{-\infty}^\infty} [u_t-u_{xx}-u^2]^2 dx dt. 
 \end{align}
\begin{align}
 \intertext{From the definition of functional derivatives we have}
 \iint \frac{\delta S}{\delta u} \psi dxdt&= \lim_{h \rightarrow 0} \frac{ S(u+h \psi)-S(u)}{h} .\\
  \intertext{Combining and taking limits gives}
 \iint \frac{\delta S}{\delta u} \psi dxdt&= 2 \iint  (\psi_t -\psi_{xx} - \psi)(u_t-u_{xx}-u)  dxdt.\\
 \intertext{Remembering that $\epsilon \xi = u_t-u_{xx}-u$ gives}
 \iint \frac{\delta S}{\delta u} \psi dxdt &= 2 \epsilon \iint  \psi_t \xi -\psi_{xx} \xi - \psi \xi   dxdt.
 \end{align}
Now using integration by parts on each $\psi$ term, as appropriate, we find
\begin{multline}
\iint \frac{\delta S}{\delta u} \psi dxdt = 2\epsilon \int \left[ \psi \xi \right]_{t=0}^{t=T} dx + 2\epsilon \int \left[ \psi \xi_x- \psi_x \xi \right]_{x=-\infty}^{x=\infty}  dt\\ 
+ 2 \epsilon \iint  -\psi \xi_t -\psi \xi_{xx} - \psi \xi   dxdt. 
\end{multline}

Because we seek to minimise $\int \int \xi^2 dx dt$, we can safely demand that $\xi,\xi_x \rightarrow 0$ as $x\rightarrow \pm \infty$ and thus  $\left[ \psi \xi_x- \psi_x \xi \right]_{-\infty}^{\infty}=0$. 
We are left with the potentially problematic $\int \left[ \psi \xi \right]_0^{T} \xi dx$ term, however, since $u$ is fixed at zero for $t=0$ and at $U_f$ for $x=0, t=T$ we know $\psi$ (our perturbation to $u$) is zero in these location. Further, we can assume that $\xi=0$ for $t=T, x \neq 0$, as any non-zero forcing at these locations would increase $S$, but have no effect on $u(0,T)$. Hence  $\int \left[ \psi \xi \right]_0^{T} \xi dx=0$, and so 

\begin{align}
\label{eq:xiEquation}
\frac{\delta S}{\delta u}= 0= -\xi_t - \xi_{xx} - \xi.
\end{align}

The above can be stated as a fourth order differential equation purely in terms of $u$, but this provides little illumination. Instead we leave it as a backwards heat equation, ready to be coupled to \ref{eq:Base_SDE_Appendix}. 

As argued previously, for our optimal solution $\xi(x,T) =0$ whenever $x \neq 0$, as non-zero values of $\xi$ in these locations increase $S(\xi)$ but have no effect on $u(0,T)$. In order for $\xi$ to have any effect on $u$ it must have non-zero total mass, and since $\int -\xi_t dx = \int \xi dx$ we know that it must have non-zero total mass at time $T$. Hence $\xi(x,T)= \alpha \delta(x)$ for some $\alpha$.

Combining the above, the minimal noise carrying $u$ from $0$ to $u(0,T)=U_f$ must solve:
 \begin{equation}
 \begin{split}
  u_t&= u_{xx} + u + \epsilon \xi, \\
  u(x,0)&=0, \\
  -\xi_t&= \xi_{xx} + \xi, \\
  \xi(x,T)&= \alpha \delta(x).
  \end{split}
\end{equation}
This recovers equation (\ref{eqn:backForwardBurst}) of the main text.

Solving directly gives

 \begin{align}
  \xi(x,t)&= \alpha e^{T-t} N(x,T-t)\\ 
  u(x,t)&= \int_0^t \alpha \epsilon e^{T+t-2 \tau} N(x,T+t-2 \tau) d\tau
  \end{align}
Where 
 \begin{equation}
  N(x,\gamma)=  \frac{1}{\sqrt{4 \pi \gamma}} \exp \left[\frac{-x^2}{4\gamma} \right] 
\end{equation}
The noise scaling $\alpha$ can be found using our boundary condition:
 \begin{align}
U_f &=  u(x,t)\\
&=\int_0^T \alpha \epsilon e^{2T-2 \tau} N(0,2T-2 \tau) d\tau\\
\label{eq:startBlockB}
&=\alpha \epsilon \int_{2T}^0 e^{\gamma} \frac{1}{\sqrt{4 \pi \gamma} } \frac{d\gamma}{-2}\\
&= \frac{\alpha \epsilon}{4 \sqrt{\pi}} \int_{0}^{2T} \gamma^{-1/2} e^{\gamma} d\gamma\\
&= \frac{\alpha \epsilon}{4} erfi(\sqrt{2T}) = \frac{\alpha \epsilon}{4} O(e^{2T})
\end{align}

In order to find the normalized action functional we find:
\begin{align}
 \label{eqn:solutionBurstLinear}
  S(\xi)&= \iint\epsilon^2 \xi^2 dx dt, \\
  &= \epsilon^2 \alpha^2 \iint e^{2T-2t} N(x,T-t) N(x,T-t) dx dt, \\ 
    &= \epsilon^2 \alpha^2 \iint e^{2T-2t} \frac{N(x,[T-t]/2)}{\sqrt{8 \pi (T-t))}} dx dt. \\
    \intertext{Remembering that $\int N(x,\gamma) dx =1$ for any $\gamma$, followed by the substitution $2(T-t)=\gamma$}
    &= \frac{\epsilon^2 \alpha^2}{4\sqrt{\pi}} \int_0^{2T} \gamma^{-1/2} e^{\gamma} d\gamma. \\
    &= \epsilon \alpha U_f = 4 U_f^2/erfi(2T) = U_f^2 O(e^{-2T})
    \label{eq:endBlockB}
  \end{align}
  
 Escape is predicted to occur when $\epsilon^{-1}=O(e^{2T})$. For all but the smallest $\epsilon$ values, this can be considered to take place in $O(1)$ time.
 
Conceptually similar calculations can be used to get from equation (\ref{eq:StartAnalysis}) to 
(\ref{eqn:backForwardBurst}) in the non-linear case.

\bibliographystyle{abbrvnat}
\bibliography{MAIN}

\end{document}